\documentclass[acmsmall,nonacm]{acmart}
\pdfoutput=1

\setcopyright{none}
\copyrightyear{2019}
\acmYear{2019}
\acmDOI{}

\acmJournal{TCPS}
\acmVolume{0}
\acmNumber{0}
\acmArticle{}
\acmMonth{0}

\usepackage{macros}

\begin{document}

\title[Safe and Private Forward-Trading Platform for Transactive Microgrids]{Safe and Private Forward-Trading Platform for\\Transactive Microgrids}


\author{Scott Eisele}
\email{scott.r.eisele@vanderbilt.edu}
\affiliation{%
  \institution{Vanderbilt University}
  \city{Nashville}
  \state{TN}
}

\author{Taha Eghtesad}
\email{teghtesad@uh.edu}
\affiliation{%
  \institution{University of Houston}
  \city{Houston}
  \state{TX}
}

\author{Keegan Campanelli }
\email{keegan.m.campanelli@vanderbilt.edu }
\affiliation{%
  \institution{Vanderbilt University}
 % \city{Nashville}
 % \state{TN}
}

\author{Prakhar Agrawal}
\email{}
\affiliation{%
  \institution{Vanderbilt University}
%  \city{Nashville}
%  \state{TN}
}

\author{Aron Laszka}
\email{alaszka@houston.edu}
\affiliation{%
  \institution{University of Houston}
%  \city{Houston}
%  \state{TX}
}

\author{Abhishek Dubey}
\affiliation{%
  \institution{Vanderbilt University}
%  \city{Nashville}
%  \state{TN}
}
\email{abhishek.dubey@vanderbilt.edu}

%\renewcommand{\shortauthors}{}

\begin{abstract}
%\iScott{can you review the abstract?}
%\iAron{Let's begin the abstract with a problem statement.}
%\platform is a market platform for peer-to-peer energy trading that satisfies the seemingly conflicting requirements of safety and privacy. It is built upon a distributed ledger platform with smart contract capabilities, which enables the core functionality of the market and ensures safety. Privacy is provided through a distributed mixing service, and safety is ensured through \textit{energy assets}\Aron{I'm not sure if the reader will be able to make sense of `energy  assets' and `group constraints' at this point.}, and checking offers against \textit{group constraints}. We provide initial implementation of the platform and underlying smart contract.  
Transactive microgrids have emerged as a transformative solution for the problems faced by distribution system operators due to an increase in the use of distributed energy resources and rapid growth in renewable energy generation. %, such as wind and solar power. 
Transactive microgrids are tightly coupled cyber and physical systems, which require resilient and robust financial markets where transactions can be submitted and cleared, while ensuring that erroneous or malicious transactions cannot destabilize the grid. In this paper, we introduce \platform, a novel decentralized platform for transactive microgrids. \platform enables participants to trade in an energy futures market, 
% which improves efficiency by increasing the likelihood of finding feasible matches for energy trades.
which improves efficiency by finding feasible matches for energy trades, reducing the load on the distribution system operator. 
\platform provides privacy to participants by anonymizing their trading activity using a distributed mixing service, while also enforcing constraints that limit trading activity based on safety requirements, such as keeping power flow below line capacity. %In the basic implementation these groups are restricted to a feeder because of the need to be ensure the net sum of energy transfers remains below the feeder safety limit. 
%In this paper we show a transformation to extend the groups to a set of multiple feeders, effectively making the group a pseudo feeder and describe the formulation to select the safety limits of the group in a way that will ensure the safety constraints of the physical feeders are still satisfied while maximizing the energy that can be traded in the system. 
We show that \platform can satisfy the seemingly conflicting requirements of efficiency, safety, and privacy, and we demonstrate its performance using simulation results.

%by anonymizing the offers made by the participant through a distributed mixing service. However, this is challenging because of the need to maintain safety which requires that the net energy transfer across a feeder is under the safety limits.

%In the paper, we describe a transformation that enables us to provide privacy and 

%and a protocol that uses per prosumer limits and per feeder limits

%effectively creates a pseudo energy 

%In the paper we make describe how we handle the  emphasis on the challenge that appears when we address safety and privacy together. The 

%in the sense that producers should not be able to create a transaction without actually producing energy, are tamper proof, and (c) are always executed within the specified time limit

\end{abstract}




\maketitle
% \section{Outline}
% \begin{itemize}
%     \item Introduction - setting up what is the role of decentralized transactive energy systems and what is this paper about. Contribution: We provide an extended protocol that ensures privacy. We explore the affect of privacy on the efficiency of the decentralized marker. Finally, we introduce the gridlabd test bed for transactive energy.
%     \item Transactive Energy Systems : A background and current status - think of this as the related research section. 
%     \item System Model and Assumptions: Explain clearly our system model and assumptions.
%     \item Privacy-Preserving Transactive Energy protcols
%     \begin{itemize}
%         \item Explain the core concepts and the workflow and the components.
%         \item explain how does the market work. How is this different than a centrally clearing market
%         \item Explain the concept of futures trading and finalization. Describe the impact of storage and how it changes the basic problem.
%         \item Explain the problem of groups and how it affects the market safety setting. Provide a few examples here.
%     \end{itemize}
%     \item Analysis of the Impact of Privacy
%     \item The Gridlab-d Testbed
%     \item Results
%     \item Discussion and conclusion
% \end{itemize}

\setlength{\marginparwidth}{1.22cm}

\section{Introduction}
\label{sec:intro}
%\Aron{Abhishek, please check the acknowledgement footnote at the bottom of this page!}

% \iflong{
% The legacy architecture of the power grid contains a number of operators,  including   an independent system operator (ISO), a transmission system operator (TSO), and a distribution system operator (DSO) hierarchically arranged across the system. The ISO is responsible for a large region and collects requirements from several DSOs and uses it to purchase energy from bulk producers.  The TSO is responsible for transferring  the energy to DSO. DSO's supply energy to distribution feeders.  

% This legacy setup is rapidly changing with the accelerating integration of renewable energy sources and battery storage solutions. \cite{keck2019impact,EIA2014, 5430489}.  
% }
% \fi

The traditional setup of the power grid is rapidly changing. Solar panel capacity is estimated to grow from 4\% in 2015 to 29\% in 2040 \cite{Randal}, and with the decreasing costs of battery technology, it is becoming increasingly feasible to support almost 99\% of total load with renewable sources, by balancing out the intermittence with batteries \cite{lehtola2019solar}. These changes are also leading to the development of a decentralized vision for the future of power-grid operations, in which local peer-to-peer energy trading within microgrids can be used to reduce the load on distribution system operators (DSO), leading to the development of transactive energy systems (TES) \cite{kok2016society,rahimi2012transactive, cox2013structured,melton2013gridwise}. 
A transactive energy system is a set of market based constructs for dynamically balancing the demand and supply across the electrical infrastructure \cite{melton2013gridwise}. In this approach, customers connected by transmission and distributions lines can participate in an open market, trading and exchanging energy locally. Customers participating in these markets are known as \textit{prosumers}. %Customers may agree on trades in advance, which allows them to schedule their transfer of power in advance. 
There are typically three phases in the operation of this market: posting offers to buy or sell energy, matching selling offers with buying offers, and synchronized energy transfer to and from the grid. 

% In practice, implementing these systems requires a distributed infrastructure comprising of smart meters, smart inverters, utility substations, utility central offices, and the transmission system operator, which has to provide the necessary computation fabric to support the interplay between the energy control and the fiscal market functions. Distribution System Operators are the custodians of these markets, while still meeting the net demand \cite{7462854}. That is, they are responsible for meeting the  residual demand not met through the local market. 
% \Aron{Here, list only things that appear in our platform. Otherwise, reviewers might wonder why neglect them.}
 
% \Aron{Here, we talk about \emph{control}, but we don't deal with control in this paper (at least not in a traditional sense that many CPS people would expect). I suggest continuing to use the market terminology from the preceding paragraph.}
% In theory, these interactions could happen in a centralized manner by communicating all relevant values to a central controller, which would compute and broadcast the ``optimal'' control settings back to each individual prosumer. However, in a centralized solution \cite{KAUR2016338}, all controllable resources would have direct communication links to the centralized controller, which would present a single point of failure. 
In theory, these interactions could happen in a centralized manner by communicating all offers to a centralized market, which would match the offers and broadcast the trades back to the individual prosumers. However, in a centralized solution  \cite{KAUR2016338}, the market presents a single point of failure. 
Building a decentralized market for transactive microgrids is challenging because the system has to satisfy a number of requirements. 
First, the market must be {\bf efficient}: the system should maximize the utilization of local supply---in meeting local demand---by matching the prosumers in the microgrid, taking advantage of their temporal flexibility in production and consumption. This requirement is crucial since effective trading is the purpose of the system; all other requirements are supporting this one. 
Second, the market must be {\bf safe}: the system must prevent negligent or malicious trading from endangering the stability or physical safety of the microgrid by rejecting trades that are unlikely to be delivered or would violate line capacity constraints. \footnote{Note that this is orthogonal to the \emph{physical enforcement of safety} which is provided by overcurrent protection units that limit the total current flowing through the microgrid.} 
Third, the market must be {\bf private}: the system should conceal information that could be used to infer the prosumers' energy usage patterns. The amount of energy produced, consumed, bought, or sold by any prosumer should be anonymous to other prosumers and limited to the monthly bill for the DSO. This is necessary because such information could be exploited, e.g., to determine when a resident is at home. Fourth, the market must be {\bf secure}: the system must ensure authenticity, data integrity, and auditability for offers, trades, and bills.  
% I REMOVED INTEGRITY FROM THE LAST POINT BECAUSE IT IS ALREADY INCLUDED IN SECURITY
Finally, the market must be {\bf resilient}: it must retain availability even if some nodes or entities (e.g., DSO) are unavailable.

%\iScott{@Abhishek, to address your comment I removed the text that was here and referred to a survey paper that mentioned them, since the references in the old text were bad. Can you review it?}
% \ad{we probably want to expand on this a bit. And specifically out of four requirements below what was satisfied and what was not}

The research community is increasingly  advocating the use of  distributed ledgers in the energy sector~\cite{andoni_blockchain_2019}. This is primarily because a distributed ledger can provide an immutable, complete, and fully auditable record of all transactions that have occurred within a system.  
%Further, it enables secure transfer of digital assets between parties through the use of decentralized code call smart contracts. 
%enables the digital representation of energy and financial assets and their secure transfer from one set of parties to another, often through the use of decentralized code call smart contracts. 
%By design, the security of this value transfer is guaranteed by the interaction protocol itself and obviates the need for trusted transaction intermediaries. The execution of smart contracts (i.e., code that captures the market logic and participants' roles) is automated and guaranteed~\cite{underwood2016blockchain,mavridou2019verisolid}. 
%Additionally, the blockchain constitutes an immutable, complete, and fully auditable record of all transactions that have occurred within the system. 
%These properties ensure market transparency, as well as the availability of a detailed market load profile and grid utilization data.  
However, there are still research gaps. 
For example, Adoni et al.~\cite{andoni_blockchain_2019} surveyed 140 applications of distributed ledgers and found that 33\% were focused on decentralized energy trading, with privacy preservation being one of the key challenges that has not been addressed. They also highlighted balancing supply to demand (stability) as another critical issue. Some works validate transactions using the DSO to ensure that the decentralized energy trades agreed between parties can actually take place.  However, doing so comes at a loss of privacy because DSO is made aware of each trade. 
%\textcolor{blue}{However, to the best of our knowledge, they do not use smart contract for addressing safety and stability, \Aron{Why is this an innovation? Just using a smart contract is not an innovation. Providing privacy and resilience with safety could be an innovation. Is that missing from prior work?} which is an innovation we address}. 
Brenzikofer et al. consider privacy mechanisms \cite{brenzikofer_quartierstrom_2019}. However, they only incentivize stability though dynamic grid tariffs, and do not actively enforce stability in the sense of limiting trading that can destablize the grid. The work presented this paper addresses the problem of privacy while ensuring the system's safety can be enforced.

%\Aron{Make it clear that we mean enforcing in the sense of limiting trading, not physical enforcement.}. They state that the market needs to account for the grid topology, but they also assume that the identity of the bid originator is private except for price and quantity, meaning that tariffs are set based on current load, and will not influence behavior until the next 15 minute interval.} 

{\bf Contributions:} In this paper, we introduce \platform\footnote{This paper is a significant extension of our previously published conference papers~\cite{Laszka17,transax,eisele2018solidworx}.}, a distributed-ledger based transactive energy system. In contrast to prior work, the key innovation of \platform is addressing the requirements of safety, efficiency, privacy, security, and resilience together. The specific contributions of this paper are as follows.

To ensure trading safety, we introduce energy production and consumption assets that represent trading allowances. Based on these assets, we enforce constraints, which can correspond to line capacity constraints, on individual prosumers as well as on groups of prosumers (e.g., prosumers that are connected to the same feeder). Note that the trading safety requirement is not difficult to satisfy by itself; it poses a challenge when combined with the privacy requirement. 
To provide privacy, we integrate a mixer that enables prosumers to anonymize their production and consumption assets within their groups, and hence trade anonymously.
We show that prosumers can remain anonymous within their groups, while the system can still enforce safety constraints (e.g., feeder line capacity) by transforming them into group constraints.
%Further, we introduce the concepts of groups which act as pseudo feeders and ensure anonymity among the prosumers that are part of the group. Through analysis, we show that prosumers can remain anonymous within the group. The groups have a transformed safety constraint that guarantees the safety across the original feeder. 
We also introduce privacy-preserving billing that enables prosumers to disclose their trades only to their smart meters, which can then perform billing without disclosing anything to the DSO other than the billed amount. 

To improve trading efficiency, we  provide prosumers with the ability to specify production and consumption offers with temporal flexibility.  We solve the trading problem as a linear program, maximizing the energy traded over a long time horizon. 
We introduce a hybrid architecture for solving this linear program, combining the trustworthiness of distributed ledgers with the efficiency of conventional computational platforms.
This hybrid architecture ensures the integrity of data and computational results---as long as majority of the ledger nodes are secure---while allowing the complex computation to be performed by a set of redundant and efficient solvers.
%The smart contract ensures that consensus on the approved transactional solution, while allowing the computation of the complex optimization problem to happen outside the smart contract.  The solution presented in this paper ensures resilience by using  selects a correct matching trade solution as long as there is one correctly functioning solver.  
The underlying communication substrate is implemented by using a distributed middleware,  called Resilient Information Architecture Platform for Smart Grid (RIAPS) \cite{eisele2017riaps,Scott2017ICCPS}.
%, which provides resilience against failure of individual services. However, we do not discuss this in detail in the paper for brevity. 

{\bf Outline:}  We describe our futures trading approach, including extensions to support safety and privacy, in Section~\ref{sec:approach}. We introduce the components of \platform in Section \ref{sec:components}, and detail its protocol in Section~\ref{sec:protocol}. We analyze how \platform meets key requirements and discuss the tradeoff between privacy and efficiency in Section~\ref{sec:analysis}. We present an integrated testbed using GridLAB-D \cite{chassin2008gridlab} and numerical results in Section~\ref{sec:experimental}. Finally, we give an overview of related work in Section~\ref{sec:related}, followed by conclusions in Section~\ref{sec:conclusion}.

  \iflong{
        \subsection{Requirements}
        \label{sec:requirements}
        % \Aron{What do we mean by `wide-scale'?}
        % To enable wide-scale microgrid-level integration, we present the following requirements that must be met by a transactive energy system. 
        To enable peer-to-peer energy trading, we present the following requirements that must be met by a transactive energy system.
        
        \begin{description}[leftmargin=0cm]
            \item [Operational Safety:] 
            % The transactive energy system must be able to prevent negligent or malicious trading from endangering the stability or physical safety of the microgrid. That is, it should reject trades that could violate line capacity constraints or are unlikely to be delivered. Each feeder is rated for a maximal power capacity.\Aron{Mention that our work is orthogonal to physical enforcement.}\footnote{Physically, the capacity limit of a feeder will be enforced using an overcurrent protection unit that limits the total current flowing through the feeder.} Therefore, it is important to ensure that local energy trade settlements involve power production and consumption which result in  power flows that never violate this safety constraint.\Aron{This is a bit repetitive here, second sentence already mentioned not violating line capacity.}
            
            The transactive energy system must be able to prevent negligent or malicious trading from endangering the stability or physical safety of the microgrid by rejecting trades that are unlikely to be delivered, or violate line or feeder capacity constraints. \footnote{This is orthogonal to the physical enforcement of safety which is provided by overcurrent protection units that limit the total current flowing through the microgrid.}

            \item [Trading Efficiency:] The transactive energy system should maximize the utilization of local supply---in meeting local demand---by matching the prosumers in the microgrid, taking advantage of the temporal flexibility in production and consumption. This requirement is crucial since effective trading is the purpose of the system; all other requirements are supporting this one.
            \item [Privacy:] 
            %Information such as the amount of energy produced, consumed, bought, or sold by any prosumer should be anonymous to both other prosumers and to the DSO. The system should limit the information that could be used to infer the prosumers' energy usage patterns, which could be exploited, e.g., to determine when a resident is at home.\Aron{Perhaps mention that DSO should get only monthly bill?}
            
            The system should limit the information that could be used to infer the prosumers' energy usage patterns. The amount of energy produced, consumed, bought, or sold by any prosumer should be anonymous to other prosumers and limited to the monthly bill for the DSO. This is necessary because such information could be exploited, e.g., to determine when a resident is at home. 
            
            \item [Security and Resilience:] The system must ensure authenticity, data integrity, and auditability for receiving and storing offers, trades, and matches; it should retain availability and system integrity even if some nodes (e.g., prosumers, solvers) are unavailable or compromised. Further, the 
            %The first part of this requirement is easily satisfied by and underlying ledger platform, if used and standard security protocols. However, there is still the requirement of network level resilience and security. Additionally, the 
            system should be able to tolerate failures in platform-level management entities.
        \end{description}

        \subsection{Contributions}

            \begin{figure*}[t]
            \centering
            \includegraphics[width=0.75\textwidth]{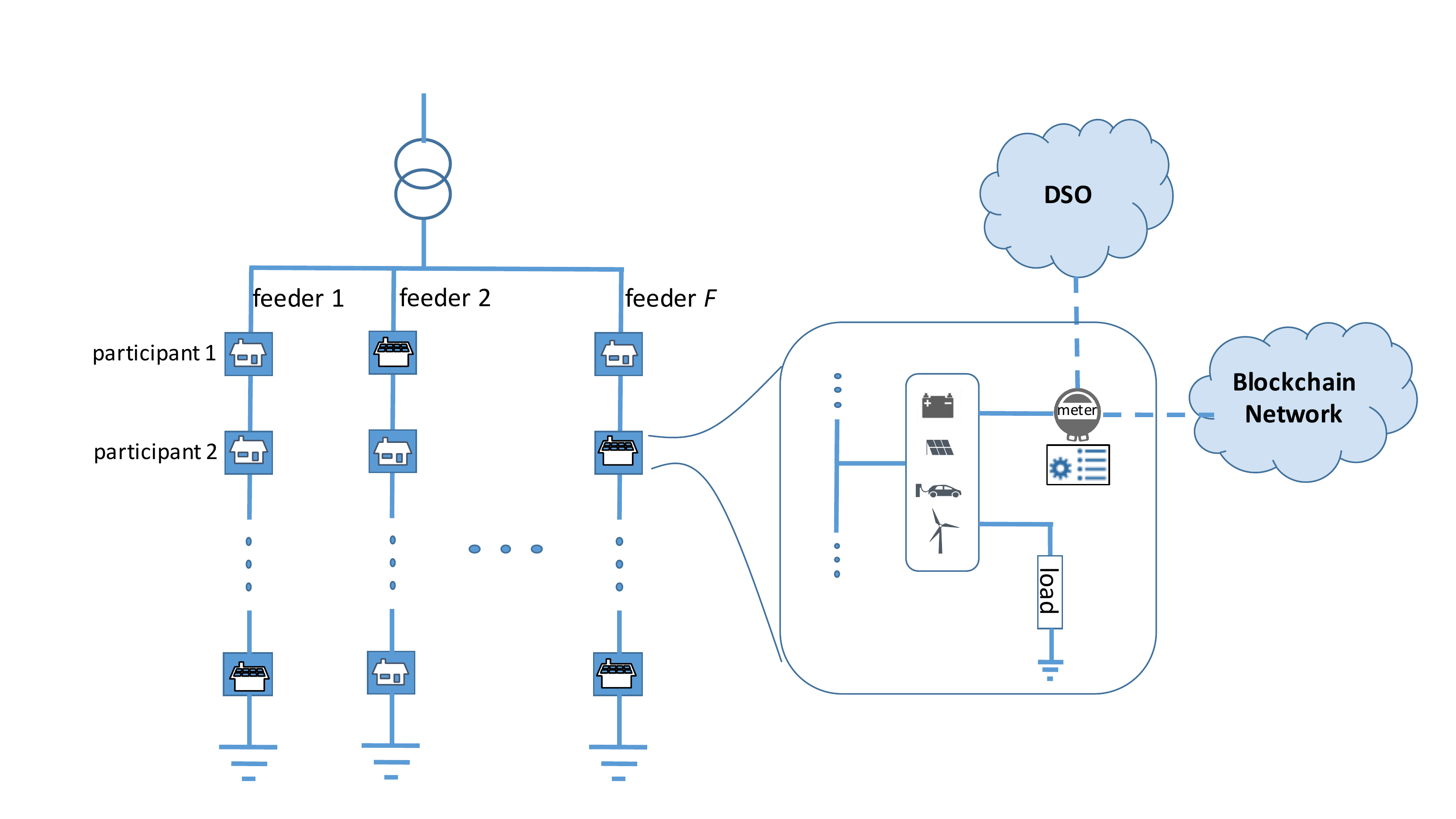}
            \vspace{-1.25em}
            \caption{We consider a multi-feeder microgrid. A feeder has number of nodes, some of whom have the capability to sell energy. Each feeder is protected by an overcurrent relay at the junction of the common bus. The inset figure shows that a node in the network has different kinds of loads, some of which can be scheduled, making it possible for a consumer to bid in advance for those loads. The smart meter ensures proper billing per node. The blockchain network is an immutable record of all transactions and is used for scheduling energy transfers between homes in the microgrid and between homes and the DSO.%and also by the Distribution System Operator (DSO) to produce a monthly bill.
            }
            \label{fig:microgrid}
            \end{figure*}

        \Aron{Since we don't have much space, we will probably need to shorten the list of contributions (also, we should make it a concise list instead of paragraphs). If we really want to save space, we could combine it with the paper outline.}
        
        This paper presents our integrated solution to the transactive energy problem  for distribution microgrids
        \iflong{(see Fig. \ref{fig:microgrid})}\fi, handling all the four requirements of operation stability, trading efficiency, prosumer privacy and security and resilience. While we have introduced different sub-features of this architecture before, this is the first paper that provides the complete overview and also provides several extensions. The initial problem formulation was introduced in \cite{transax,eisele2018solidworx} focusing on the base operational safety problem. We had introduced the concepts of privacy architecture earlier in \cite{Laszka17}, however, it was not integrated into the operational protocol of \cite{transax} due to the conflicting problem of privacy and safety. The conflict arises from need of anonymization for privacy and the requirement of being able to monitor the energy exchanges and ensuring that they do not violate the feeder safety limits.  
        
        %For example,
        \platform uses blockchain to implement the transactive platform. Although disintermediation of trust is widely regarded as the primary feature of blockchain-based transaction systems \cite{SpectrumBC}, their use in TES is appealing also because they elegantly integrate the ability to immutably record the ownership and transfer of assets, with essential distributed computing services such as Byzantine fault-tolerant consensus on the ledger state as well as event chronology.  The ability to establish consensus on state and timing is important in the context of TES since these systems are envisioned to involve the participation of self-interested parties, interacting with one another via a distributed computing platform that executes  transaction management However, this also leads to the problem of privacy as the records in blockchain can be attributed to the  prosumers. 
        
        We solve this challenge by introducing a mixer that enables prosumers to anonymize their production / consumption assets, and hence post offers and trade anonymously. The main contribution here is to provide privacy while still enforcing the safety constraints. Thus, we introduce the concepts of groups which act as pseudo feeders and ensure anonymity between the prosumers that are part of the group. Through analysis, we show that prosumers can remain anonymous within the group. The groups have a transformed safety constraint that guarantees the safety across the original feeder. \st{Later we present heuristics to select the groups and show how the group formation, and subsequently the level of anonymity affects the trading efficiency}. We also provide privacy-preserving billing by enabling prosumers to disclose their trades to their smart meters in a trustworthy way, thereby enabling smart meters to perform billing for transactive microgrids. This is important because the DSO will only receive a bill (e.g., monthly) instead of fine-grained information. Note that smart meters can perform time-variant billing (e.g., higher energy prices from DSO at peak hours). 
        
        To ensure operational safety, we introduce energy production / consumption assets that represent trading allowances. Further, we enforce constraints, which can correspond to line capacity constraints, on individual prosumers and groups using the assets and the smart contract. Note that this requirement is not difficulty to satisfy by itself; it becomes tricky when combined with privacy. We discuss this later in the text. We also show that via analysis that no trades will be recorded that could violate any of the constraints.
        
        To improve the trading efficiency, we  provide prosumers with the ability to specify production and consumption offers with temporal flexibility.  We solve the trading problem as a linear program, maximizing the energy traded during a time quantum. We introduce a hybrid architecture, combining the trustworthiness of the ledger with the efficiency of conventional computational platforms, for solving the linear program.  This hybrid architecture ensures that the key data of the platform remains secure as long as most of the ledger nodes are secure. The smart contract ensures that consensus on the approved transactional solution, while allowing the computation of the complex optimization problem to happen outside the smart contract\footnote{The smart contract computations are expensive and we use them only for validation of the solution rather than for computing the solution}. Through analysis, we show that the energy trading can reduce the load on the power grid and that temporal flexibility can significantly increase trading volume.
         
         The solution presented in this paper ensures resilience by using  the distributed ledger and a protocol that can tolerate failures. The underlying communication substrate is implemented by using a distributed middleware  called Resilient Information Architecture Platform for Smart Grid (RIAPS) \cite{eisele2017riaps,Scott2017ICCPS} that provides resilience against failure of individual services. Further, use of RIAPS solves the device interoperability issues \cite{deviceaccess} that can be caused by the different devices in the distribution microgrid: smart meters, smart inverters, etc.  Through analysis, we show that the platform can function even if some components are unavailable or compromised. We further show that the smart contract will select a correct matching as long as there is one correctly functioning solver.  
         }
\fi

{\iflong
\section{Background Concepts}
\label{sec:background}

To explain the architectural concepts we first need to provide an overview of the basic concepts and how we use them in \platform.

\subsection{Resilient Information Architecture for Smart Grid}
\label{sec:riaps}

% \Aron{Can we shorten this since it is already published?}
RIAPS is an \textit{open application platform} for smart grid that distributes intelligence and control capability to local endpoints to reduce total network traffic, avoid latency, and decrease dependency on multiple devices and communication interfaces, thereby enhancing reliability. %The platform  can host several applications running concurrently, enabling agents to communicate with each other. It focuses on certain grid issues, such as state estimation, remedial action schemes, and load shedding. 
It also provides platform services to power systems applications running on remote nodes  \cite{Volgyesi2017Time}, including:
\begin{enumerate*}
    \item  a resource-management framework to control use of computational resources, \item  a fault-management framework to detect and mitigate faults in all layers of the system, \item a security framework to protect confidentiality, integrity, and availability of a system under cyber-attacks, \item a fault tolerant time synchronization service,  \item a discovery framework to establish the network of interacting actors for an application, and \item  a deployment and management framework for administration of the distributed applications from a control room.
\end{enumerate*} 
In \platform, RIAPS is used as the base middleware and application management substrate, allowing all actors to communicate and the protocol discussed later in this paper to be implemented. As a note, actor interactions are supplemented with interactions with the distributed ledger and the smart contract. We will discuss these interactions in detail in the protocol section. For more details on RIAPS see our paper \cite{eisele2017riaps}.

\subsection{Distributed Ledgers and Smart Contract}

Distributed ledgers extend the core principles of Blockchain technology to beyond that of cryptocurrency.  These ledgers refer to distributed databases that are maintained by byzantine consensus between parties i.e., more than 50\% eventually agrees to the same reality. The data in the ledger are maintained in a chain of immutable blocks that are cryptographically sealed. Nodes in the network simultaneously reconcile their copies of the data through consensus to achieve a shared truth, so that data in the shared ledger can be verified and is tamper-aware. Data is added to the ledger via transactions and all transactions submitted to the ledger are associated with an account, which is typically not anonymous. Critical to the success of DLT is the consensus process that eventually orders all valid transactions.

In \platform we use a ledger to permanently store transactions.  The immutability of actions is crucial for providing safety and security, \emph{i.e.}, after a transaction has been recorded, it cannot be modified or removed from the ledger.  The ledger is distributed to enhance fault tolerance. Since a distributed ledger is maintained by multiple nodes, nodes must reach a consensus on which transactions are valid and stored on the ledger.  This consensus must be reached both quickly and reliably, even in the presence of erroneous or malicious (\emph{e.g.}, compromised) ledger nodes. We make no assumptions about the particulars of the consensus algorithm.  In practice, a distributed ledger can be implemented using, \emph{e.g.}, \emph{blockchains} with proof-of-stake consensus or a practical Byzantine fault tolerance algorithm~\cite{castro1999practical}.

More recent distributed ledgers like Ethereum \cite{EthereumBook}, which we use in this work enable creation of smart contracts, that is a deterministic code which reads input from the ledger and records the output generated by the execution of the code back in the ledger. The consensus mechanism ensures that all participants of the system agree to the outcome of the execution. Since smart contracts can perform any Turing complete deterministic computation, they can be used to write complicated decentralized applications. We use such a smart contract in the \platform. 

Note that a big concern with writing decentralized applications with smart contract is the cost of the resources required to ensure consensus, which in effect translates to an increased cost of computation. Further, several available smart contract implementations are constrained in their ability to represent floating point variables, which limit their direct use in solving problems like optimization. This is one of the reason we do not use smart contract to search for a matching transactive energy solution. Rather, we only use it for validation of a given transactive energy solution.

\subsubsection{Microgrid}
We consider a microgrid with a set of feeders. A feeder has a fixed set of nodes, each representing a residential load or a combination of load and distributed energy resources, such as rooftop solar and batteries, as shown in Figure~\ref{fig:microgrid}. Each node is associated with a participant in the local peer-to-peer energy trading market and each participant is independent and has control over their energy utilization. The participants can predict their future production and consumption based on historical data and anticipated utilization.

\subsubsection{Distribution System Operator}
We assume the existence of a distribution system operator (DSO) that also participates in the market, and may use the market to incentivize timed energy production within the microgrid to aid in grid stabilization and promotion of related ancillary services \cite{7462854} through updates to the price policy. In addition, the DSO supplies residual demand not met through the local market. The participants settle trades in advance using automated matching which allows them to schedule their transfer of power into the local distribution system. We also assume the presence of a secondary controller that balances voltage and frequency in the microgrid. We describe such a controller in~\cite{haoriaps17}.

\subsubsection{Smart Meters}
To measure the prosumers' energy production and consumption, a smart meter must be deployed at each prosumer.  In practice, these smart meters must be tamper resistant to prevent prosumers from ``stealing electricity'' by tampering with their meters.  After a smart meter has measured the net amount of energy consumed by the prosumer in some time interval, it can send this information to the DSO for billing purposes.

\subsection{Mixing Protocols}
\label{sec:mixing}

The use of blockchains in building a transactive energy platform is  appealing also because they elegantly integrate the ability to immutably record the ownership and transfer of assets, with essential distributed computing services such as Byzantine fault-tolerant consensus on the ledger state as well as event chronology.  The ability to establish consensus on state and timing is important in the context of TES since these systems are envisioned to involve the participation of self-interested parties, interacting with one another via a distributed computing platform that executes  transaction management However, this also leads to the problem of privacy as the records in blockchain can be attributed to the  prosumers.

The earliest approach to solve the privacy problem in Blockchains was mixing. The key concept in mixing is to hide the linkage between inputs and outputs of transactions by combining them with the information of other transactions.  In \platform,  we use coinshuffle \cite{ruffing2014coinshuffle}, and include a short description of how it works in our appendix. %and describe it below.

\fi}

% \section{Related Work}
% \label{sec:related}

\section{The Energy Trading Approach} 
\label{sec:approach}
% In this section, we formalize our approach to \emph{efficient energy trading}. \change{, including offers, and how they can be matched and anonymized}{}. We then extend the formalization to \Aron{``to satisfy requirements for safety, privacy, and resilience.'' (or something like this)}make it more practical, and support safe anonymization at more than just the feeder level. 
In this section, we formalize our approach to \emph{efficient energy trading} in \platform. We then extend the formalization to support safety, privacy, and resilience. 
%We also modify it to make finding the optimal solution more practical. 

\iflong{Recall that we }\else{We }\fi assume the distribution network infrastructure to be a  microgrid with a set of feeders. A feeder has a fixed set of nodes, each representing a prosumer, which is a combination of load and distributed energy resources, such as rooftop solar panels and batteries. We assume that the prosumers can predict their future production and consumption based on historical data and anticipated utilization. The prosumers submit energy offers based on their predictions via automated agents that act on behalf of residents (i.e., residents do not need to trade manually). The predictions do not need to be perfect because we assume the existence of a distribution system operator (DSO), which also participates in the market and can supply residual demand not met through the local market. The DSO may use the market to incentivize timed energy production within the microgrid to aid in grid stabilization and in the promotion of related ancillary services \cite{7462854} through updates to the price policy. To measure the prosumers' energy production and consumption, a smart meter must be deployed at each prosumer.  In practice, these smart meters must be tamper resistant to prevent prosumers from ``stealing electricity''.  After a smart meter has measured the net amount of energy consumed by the prosumer in some time interval, it can send the relevant information to the DSO for billing purposes. 

Our goal is to find an optimal match between energy production and consumption offers, which we refer as the \emph{energy trading problem}. Each offer is associated with an identity that belongs to the prosumer that posted the offer. We refer to these identities as \textit{accounts}, and prosumers may generate any number of them.

\begin{table}
\footnotesize
\caption{List of Symbols}
\vspace{-1.2em}
\centering
\label{tab:symbols}
\renewcommand*{\arraystretch}{1}
\begin{tabular}{|l|p{11cm}|}
\hline
Symbol & Description \\
\hline
\multicolumn{2}{|c|}{Microgrid} \\
\hline
$\calF$, $\calU$ & set of feeders and prosumers, resp. \\
\rowcolor{TableRowGray} $C_f^{ext}$, $C_f^{int}$ & maximum allowed net and total, resp., power consumption or production in feeder $f \in \calF$ \\
$EPL_{u}$, $ECL_{u}$ & maximum allowed production and consumption, resp., of prosumer $u$ \\
\rowcolor{TableRowGray} $C_g^{ext}$, $C_g^{int}$ & maximum allowed net and total, resp., power consumption or production in group $g \in \calG$ \\
$EPA$, $ECA$ & asset granting permission to produce or consume, resp., a unit of energy \\
\rowcolor{TableRowGray} $\Delta$ & length of each time interval \\
$T_{clear}$ & minimum number of time intervals between the finalization and notification of a trade \\
%\rowcolor{TableRowGray} $t_{c}$ & current interval \\
%$t_f$ & next interval to be finalized \\ % $t_{f} = t_{c} + T_{clear} + 1$. $t_{c}+T_{clear}$ has been finalized.\\
\rowcolor{TableRowGray} $E_{u}^{t}$ & energy transferred by prosumer $u$ in interval $t$ \\
$t_f$ & next interval to be finalized $t+ 1 + T_{clear}$ \\

\hline
\multicolumn{2}{|c|}{Offers} \\
\hline
$\calS_f$, $\calB_f$ & set of selling and buying offers, resp., from feeder $f \in \calF$ \\
\rowcolor{TableRowGray} $\calS$, $\calB$ & set of all selling and buying offers, resp. \\
$\calS^{(t)}$, $\calB^{(t)}$ & set of all selling and buying offers, resp., submitted by the end of time interval $t$ \\
\rowcolor{TableRowGray}$A_s$, $A_b$ & account that posted offers $s \in \calS$ and $b \in \calB$, resp. \\
$E_s$, $E_b$ & amount of energy to be sold or bought, resp., by offers $s \in \calS$ and $b \in \calB$ \\
\rowcolor{TableRowGray} $I_s$, $I_b$ & time intervals in which energy could be provided or consumed by offers $s \in \calS$ and $b \in \calB$, resp. \\
$R_s$, $R_b$ & reservation prices of offers $s \in \calS$ and $b \in \calB$, resp. \\
\rowcolor{TableRowGray} $\calM(s)$, $\calM(b)$ & set of offers that are matchable with offers $s \in \calS$ and~$b \in \calB$, resp. \\
\hline
\multicolumn{2}{|c|}{Solution} \\
\hline
$p_{s,b,t}$ & amount of energy that should be provided by $s$ to $b$ in interval $t$ \\
\rowcolor{TableRowGray} $\pi_{s,b,t}$ & unit price for the energy provided by $s$ to $b$ in interval~$t$ \\
$\textit{Feasible}(\calS, \calB)$ & set of feasible solutions given sets of selling and buying offers $\calS$ and $\calB$ \\
\rowcolor{TableRowGray} $\hat{p}_{s,b,t}$, $\hat{\pi}_{s,b,t}$ & finalized trade values \\
\hline
\multicolumn{2}{|c|}{Implementation Parameters} \\
\hline
% $T_{predict}$ & prediction horizon (number of intervals beyond $t_{f}$ for which prosumers predict and post offers) \\
%$\Delta_s$ & periodicity of solver that matches offers \\

% \rowcolor{TableRowGray}  $T_{h}$ &  number of time intervals beyond the next interval to be finalized $t_f$ that are considered non-zero by the solver. All trade offers beyond $t_{f}+T_{h}$ are considered 0.\\
\rowcolor{TableRowGray}  $T_{h}$ &  solve horizon (offers beyond horizon $t_{f}+T_{h}$ are not considered by solver) \\

%$start$ & simulation start interval \\
%\rowcolor{TableRowGray} $end$ & simulation end interval \\
 $\hat{\Delta}$ &  length of the time step used for simulating the real interval of length $\Delta$ \\
\hline
\end{tabular}
\end{table}

\subsection{Formalization}
\label{sec:formalization}
Let $\calF$ denote the set of feeders. On each feeder, there is a set of prosumers, who can make offers to buy and sell energy. We assume that time is divided into intervals of fixed length~$\Delta$, and we refer to the $t$-th interval simply as time interval $t$. For a list of symbols used in the paper, see Table~\ref{tab:symbols}.
% \Aron{At some point, mention that we assume automated agents to act on behalf of residents (we do not assume people to directly trade); of course, this has nothing to do with the design of the platform.}

For feeder $f \in \calF$, we let $\calS_f$ and $\calB_f$ denote the set of selling and buying offers posted by prosumers in feeder $f$, respectively.\footnote{To include the DSO in the formulation, we assign it to a ``dummy'' feeder.} 
A selling offer $s \in \calS_f$ is a tuple $\left(A_s, E_s, I_s, R_s\right)$, where
%\begin{itemize}[topsep=0pt, leftmargin=1em]
%\item 
$A_s$ is the account that posted the offer, 
%\item 
$E_s$ is the amount of energy to be sold,
%\item 
$I_s$ is the set of time intervals in which the energy could be provided,
%\item 
$R_s$ is the reservation price, i.e., lowest unit price for which the prosumer is willing to sell energy. 
%\end{itemize}
Similarly, a buying offer $b \in \calB_f$ is a tuple $\left(A_b, E_b, I_b, R_b\right)$, where the values pertain to consuming/buying energy instead of producing/selling, and $R_b$ is the highest price that the prosumer is willing to pay. For convenience, we also let $\calS$ and $\calB$ denote the set of all buying and selling offers (i.e., we let $\calS = \cup_{f \in \calF} \calS_f$ and $\calB = \cup_{f \in \calF} \calB_f$).

We say that a pair of selling and buying offers $s \in \calS$ and $b \in \calB$ is \emph{matchable} if
\begin{align}
R_s \leq R_b ~~~~ \textnormal{ and }
~~~~ I_s \cap I_b \neq \emptyset .
\end{align}
In other words, a pair of offers is matchable if there exists a price that both prosumers would accept and a time interval in which the seller and buyer could provide and consume energy.
For a given selling offer $s \in \calS$,
we let the set of buying offers that are matchable with $s$ be denoted by $\calM(s)$.
Similarly, we let the set of selling offers that are matchable with a buying offer $b$ be denoted by~$\calM(b)$.

% \Aron{Perhaps mention what the `energy trading problem' is, e.g., say at the beginning of this subsection that we want to find an optimal match between production and consumption, which we call energy trading problem.}
A solution to the  energy trading problem is a pair of vectors $(\vp, \vpi)$, where
%\begin{itemize}[topsep=0pt, leftmargin=1em]
%\item 
$p_{s,b,t}$ is a non-negative amount of energy that should be provided by offer $s \in \calS$ and consumed by offer $b \in \calM(s)$ in time interval $t \in I_s \cap I_b$~\footnote{We require the both seller and buyer to produce a constant level of power during the time interval. This can be achieved by smart inverters.};
%\item 
and $\pi_{s,b,t}$ is the unit price for the energy provided by offer $s \in \calS$ to offer $b \in \calM(s)$ in time interval $t \in I_s \cap I_b$.
%\end{itemize}

A pair of vectors $(\vp, \vpi)$ is a feasible solution to the energy trading problem if it satisfies the following two constraints.
%\begin{itemize}[topsep=0pt, leftmargin=1em]
%\item 
First, the amount of energy sold or bought from each offer is at most the amount of energy offered:
    \begin{align}
    \forall s \in \calS: ~ \sum_{b \in \calM(s)} \sum_{t \in I(s,b)} p_{s,b,t} \cdot \Delta \leq E_s
    \hspace{.5cm}
    \textnormal{and}% \label{eq:constrEnergyProd},
    \hspace{.5cm}
    \forall b \in \calB: ~ \sum_{s \in \calM(b)} \sum_{t \in I(s,b)} p_{s,b,t} \cdot \Delta \leq E_b \label{eq:constrEnergy} 
\end{align}
%
%\item 
Second, the unit prices are between the reservation prices of the seller and buyer:
\begin{align}
\forall s \in \calS, b \in \calM(s), t \in I(s,b): ~ R_s \leq \pi_{s,b,t} \leq R_b \label{eq:constrPrice} 
\end{align}
%\end{itemize}

The objective of the energy trading problem is to maximize the amount of energy traded. The rationale behind this objective is maximizing the load reduction on the bulk power grid. Formally, an optimal solution to the energy trading problem is
\begin{align}
\label{optimalproblem}
\max_{(\vp,\vpi) \, \in \textit{ Feasible}(\calS, \calB)} ~
\sum_{s \in \calS} \sum_{b \in \calM(s)} \sum_{t \in I(s,b)} p_{s,b,t} \,
\end{align}
where $\textit{Feasible}(\calS, \calB)$ is the set of feasible solutions given selling and buying offers $\calS$ and $\calB$ (i.e., set of solutions satisfying Equations~\eqref{eq:constrEnergy} and \eqref{eq:constrPrice} with $\calS$ and $\calB$).

The above formulation ensures feasibility, which takes the reservation prices into account. However, we do not address how to set the clearing prices in this paper. Clearing prices could be set using an existing approach, e.g., double auction; we discuss how this could be done in Section \ref{sec:future}.
% We discuss some ways this could be done in future work. 
% \Taha{Is it already included in first paragraph of Section~\ref{sec:future}? I think we need to elaborate more.}\Aron{+1}\Scott{We talk about it some in \ref{sec:auction}. I took it out of \ref{sec:future}, if you want to say something else about it, then add it.}

\iflong{
    \begin{figure}
        \centering
        % \begin{tikzpicture}[scale=0.50, every node/.style={scale=.75}]
%     % \draw[step=1cm,gray,very thin] (-3,-3) grid (3,6);
%     \draw[thick] (0,0) -- (0,4);

%     \draw[thick] (0,1) -- (3,1) node[text width=2cm, label, xshift=1cm]{Feeder 3 $C_f=30$};
%     \draw[thick] (0,2) -- (-2.75,2) node[text width=2cm, label, xshift=-.3cm] {Feeder 1 $C_f=10$};
%     \draw[thick] (0,3) -- (3,3) node[text width=2cm, label, xshift=1cm]{Feeder 2 $C_f=20$};
    
%     \draw[thick] (1,1) -- (1,.5) node {$s=25$} ;
%     \draw[thick] (2,1) -- (2,1.5) node {$s=23$};
    
%     \draw[thick] (-2,2) -- (-2,1.5) node {$s=5$} ;
%     \draw[thick] (-1,2) -- (-1,2.5) node {$s=8$} ;
    
%     \draw[thick] (1,3) -- (1,2.5) node {$s=15$} ;
%     \draw[thick] (2,3) -- (2,3.5) node {$s=18$} ;
    
% \end{tikzpicture}

\begin{tikzpicture}[
scale=.75,x=1cm,y=1cm,every text node part/.style={align=center},
every node/.style={scale=.75},
relay/.style={rectangle, fill=black, minimum width=.05cm, minimum height=.5cm},
]

    \node(ls) at (0,0){};
    \node(le) at (0,4){};
    
    \node(f1s) at (0,2) {};
    \node(f1e) at (-3,2) {Feeder 1 \\ $C_{f}^{int}=10$};
    
    \node(f2s) at (0,3) {};
    \node(f2e) at (3,3) {Feeder 2 \\ $C_{f}^{int}=20$};
    
    \node(f3s) at (0,1) {};
    \node(f3e) at (3,1) {Feeder 3 \\ $C_{f}^{int}=30$};
    
    \node(p1) at (-1.5,1.5) {$10$};
    \node(p2) at (-1,2.5) {$10$};
    \node(p3) at (1,2.5) {$20$};
    \node(p4) at (1.5,3.5) {$20$};
    \node(p5) at (1,0.5) {$30$};
    \node(p6) at (1.5,1.5) {$30$};
    
    \draw[](ls.center) -- (le.center);
    \draw[](f1s.center) -- (f1e);
    \draw[](f2s.center) -- (f2e);
    \draw[](f3s.center) -- (f3e);

    \draw[]([shift={(-1.5,0)}]f1s.center) -| (p1);
    \draw[]([shift={(-1,0)}]f1s.center) -| (p2);
    
    \draw[]([shift={(1,0)}]f2s.center) -| (p3);
    \draw[]([shift={(1.5,0)}]f2s.center) -| (p4);
    
    \draw[]([shift={(1,0)}]f3s.center) -| (p5); \draw[]([shift={(1.5,0)}]f3s.center) -| (p6);

    \node[relay,label={$C_{f_{1}}^{ext}$}](r1) at  (.4, 3){};
    \node[relay,label={$C_{f_{2}}^{ext}$}](r2) at (-.4, 2){};
    \node[relay,label={$C_{f_{3}}^{ext}$}](r3) at  (.4, 1){};

\end{tikzpicture}
        \caption{Feeder diagram for cost-efficiency analysis. The lines below a feeder represent producers and those above are consumers. The unlabeled numbers below and above each feeder line are the sums of the energy production and consumption capacities respectively of the prosumers on each of the feeders.}
        \label{fig:feeder-group}
    \end{figure}
}\fi

\subsection{Safety Extensions}
\label{sec:extend_safety}

To ensure the safety of the microgrid, we introduce additional constraints on the solution to the energy trading problem. 
Each prosumer $u$ has independent production and consumption limits, which are denoted by $ECL_{u}$ and $EPL_{u}$, respectively. 
Further, each feeder $f \in \calF$, has a transformer for incoming power, which has a capacity rating.
We let $C_f^{ext}$ denote the capacity of the transformer of feeder $f$.
% maximum amount of power that is allowed to flow into or out of feeder $f$ at any moment. 
Similarly, the distribution lines and transformers for transferring power within the feeder have capacity ratings as well.
We let $C_f^{int}$ denote the maximum amount of power that is allowed to be consumed or produced within the feeder at any moment.\footnote{In other words, limit $C_f^{ext}$ is imposed on the net production and net consumption of all prosumers in feeder $f$, while limit $C_f^{int}$ is imposed on the total production and consumption of prosumers in feeder $f$.} These constraints are physically enforced by the over-current relays of the circuit breakers and feeders. 

Now we generalize and introduce the notion of groups. We note that groups can correspond to feeders and support the constraints that we introduced in the previous paragraphs. They allow us to support physical layouts other than strictly feeders, and it will be useful for privacy later. We define a \emph{group} $g$ to be a set of feeders (i.e., $g \subseteq \calF$). We let $\calG$ be the set of all groups, and for each group $g \in \calG$, we introduce group safety limits $C_{g}^{int}$ and $C_{g}^{ext}$, which are analogous to feeder limits. A solution is safe if it satisfies the following three constraints. %\begin{itemize}[topsep=0pt, leftmargin=1em]
%\item 
First, the amount of power flowing into or out of a prosumer is within the production and consumption limits in all time intervals: 

% \iAron{Firstly, prosumers post with accounts, not with their own identifiers. Secondly, $\calS_{P}$ and $\calB_{P}$ are undefined. This could be fixed by using $A_s$ and $A_b$: just consider all accounts $A$ that belong to the prosumer, and set of all offers $s \in \calS$ that were posted by $A$ (i.e., $A_s = A$).
% }
% \iScott{I'm not sure what that would look like. }

% \begin{align}
%         \forall u,t:  \sum_{A \in u} \sum_{\substack{s \in \calS:\\A_s = A}}\sum_{b\in \calB}p_{s,b,t} \leq EPL_{u}
%         \hspace{0.5cm}
%         \textnormal{and}
%         \hspace{0.5cm}
%         \forall u,t: \sum_{A \in u} \sum_{b\in \calB}\sum_{s\in \calS}p_{s,b,t} \leq ECL_{u} 
%         \label{eq:EL}
%     \end{align}  
\begin{align}
        \forall u \in \calU, t:  \sum_{s \in \calS_u}\sum_{b\in \calB} p_{s,b,t} \leq EPL_{u}
        \hspace{0.5cm}
        \textnormal{and}
        \hspace{0.5cm}
        \forall u \in \calU, t: \sum_{b\in \calB_u}\sum_{s\in \calS}p_{s,b,t} \leq ECL_{u} 
        \label{eq:EL}
    \end{align}  
where $\calS_u$ and $\calB_u$ are the sets buying and selling offers posted by accounts owned by prosumer $u$.
    
% \iScott{Maybe something like this? Where $O(u,A)$ means $A$ owned by $u$?}
% \begin{align}
%     \forall A \bigg( O(u,A) \land s\in \calS(A_{s}=A) \bigg) \sum_{s}\sum_{b\in \calB}p_{s,b,t} \leq EPL_{u}
% \end{align}

%\Aron{do groups really need to be sets of feeders? Why don't we define them as sets of prosumers?}
%\iScott{They could be sets of prosumers, the analysis done by the undergrad didn't take that into account though. We have deviated from that so maybe it would work with just prosumers but we'd have to check. Is there a critical issue sticking with this for now?}

Second, the amount of energy consumed and produced within each group is below the safety limit in all time intervals:
    \begin{align}
    \forall g \in \calG, t: ~ \max\left\{\sum_{b \in \calB_g} \sum_{s \in \calS} p_{s,b,t},\sum_{s \in \calS_g} \sum_{b \in \calB} p_{s,b,t}\right\} \leq C_g^{int} 
    \label{eq:constrInt}
    \end{align}
    %\Aron{Briefly explain what each term means?}
    \iflong{
    For example, for feeder 1 in Fig. \ref{fig:feeder-group}, this constraint means that at most $10$ units can be safely transferred even if the total production capacity is greater than $10$. Otherwise, the over-current relay that protects the feeder will trip.}
    \else{This means that the sum of all the buying and the selling trades cannot exceed the safety limit. }

%\item 
Third, the amount of power flowing into or out of each group is within the safety limit in all time intervals:
    \begin{align}
    \forall g \in \calG, t: 
     -C_g^{ext} \leq \left( \sum_{s \in \calS_g} \sum_{b \in \calB} p_{s,b,t} \right) - \left( \sum_{b \in \calB_g} \sum_{s \in \calS} p_{s,b,t} \right) \leq C_g^{ext} \label{eq:constrExt}
    % \forall f \in & \, \calF, t: 
    % & \left( \sum_{s \in \calS_f} \sum_{b \in \calB} p_{s,b,t} \right) - \left( \sum_{b \in \calB_f} \sum_{s \in \calS} p_{s,b,t} \right) \geq -C_f^{ext} \label{eq:constrExtCons} 
    \end{align}
    \iflong{Of the expression in the middle of the inequality, the first term is the total amount of energy sold by the group, and the second term is the total amount of energy bought by the group. In Fig. \ref{fig:feeder-group}, this means that if a prosumer in group 1 sells $10$ units to a prosumer in group 2 and a distinct prosumer in group 2 sells 10 units to a distinct prosumer in group 1 it is as if the two prosumers in group 1 traded with each other and the two in group 2 traded with each other. Basically, any trades that occurred within the group, or could have, do not end up counting towards the $C_{g}^{ext}$ limit. }
    \else{Of the expression in the middle of the inequality, the first term is the total amount of energy sold by the group, and the second term is the total amount of energy bought by the group. Any trades that occurred within the group, or could have, do not end up counting towards the $C_{g}^{ext}$ limit.}
    Note that expression compared to $C_{g}^{int}$ in Equation \eqref{eq:constrInt} is the largest of the buying and selling matches, whereas the expression compared to $C_{g}^{ext}$ in Equation \eqref{eq:constrExt} is the difference between the buying and selling matches; this means that the value compared against $C_{g}^{int}$ will always be larger than the value compared against $C_{g}^{ext}$. Thus if  $C_{g}^{int} < C_{g}^{ext}$ then the external limit will never trip, and so we only need to consider $C_{g}^{ext} \leq C_{g}^{int}$. Verifying if a solution is safe in this case is simple, the offers in the solution need only be associated with the group they came from and checked against the~constraints. 
%\end{itemize}

% To check these constraints each offer must be associated with a group. We achieve this by associating each offer with an identity that belongs to the participant that posted the offer, and each identity is associated with a group. We refer to these identities as \textit{accounts}. 

% \begin{new}
\subsection{Privacy Extensions} 
\label{sec:privacy}

To protect prosumers' privacy, we let them use anonymous accounts when posting offers. By generating new anonymous accounts, a prosumer can prevent others from linking the anonymous accounts to its actual identity, thereby keeping its trading activities private. 
However, anonymous accounts pose a threat to safety. Since the energy trading formalization with safety extension (see Equations \eqref{eq:EL} - \eqref{eq:constrExt}) discussed earlier requires the offers to be associated with the prosumer to enforce prosumer-level constraints and with the group from which they originated in order to be able to enforce group-level safety constraints. Without these associations prosumers, can generate any number of anonymous accounts, and post selling and buying offers for large amounts of energy without any intention of delivering and without facing any repercussions. A malicious or faulty prosumer could easily destabilize the grid with this form of reckless trading. Consequently, the amount of energy that may be traded by anonymous accounts belonging to a prosumer must be~limited. 

To enforce the prosumer-level constraints we introduce the concept of energy production and consumption assets, which allows us to disassociate the limiting of assets from the anonymity of offers. First, an \emph{energy production asset} (EPA) is tuple ($E_{\textnormal{\textit{EPA}}}, I_{\textnormal{\textit{EPA}}}, G_{\textnormal{\textit{EPA}}}$), where

\begin{itemize}[noitemsep,topsep=-\parskip]
    \item $E_{\textnormal{\textit{EPA}}}$ is the permission to sell a specific non-negative amount of energy to be produced,
    \item $I_{\textnormal{\textit{EPA}}}$ is the set of intervals for which the asset is valid, and
    \item $G_{\textnormal{\textit{EPA}}}$ is the group that the asset is associated with.
\end{itemize}

Second an \emph{energy consumption asset} (ECA) represents a permission to buy a specific amount of energy and is defined by the same fields.  For this asset, however, the fields define energy consumption instead of production. Each prosumer $u$ is only permitted to withdraw assets up to the limits $EPL_{u}$ and $ECL_{u}$ into a non-anonymous account.  

These assets can be moved to anonymous accounts in an untraceable way, but still retain their group association and the sum of the assets remains constant. Production assets are required to post a selling offer, and consumption assets are required to post a buying offer. For the offer to be valid, the account posting the offer must have assets that cover the amount and intervals offered. When a trade is finalized the assets are exchanged. We will provide more details on how they fit into the trading approach in Section~\ref{sec:protocol}.

%When traded, the $ECA$ ($EPA$) assets are exchanged for $EPA$ ($ECA$) assets which are also tagged so the smart meter will now they were part of a trade and not simply unused assets, so we denote assets that have been traded for as $ECA'$ and $EPA'$. 

To enforce group-level safety we only provide group-level anonymity, meaning that an offer can be traced back to its group of origin, but not to the individual prosumer within the group. When forming a group the safety constraints will need to be set appropriately. We will discuss how they should be set and the associated energy trading capacity costs in Section \ref{sec:tradeoff}.

\subsection{Iterative Solutions Extension}
\label{sec:extendedProblem}

% \Aron{I'm not entirely sure about the section title.}
% \Scott{How about this one?, or \textit{Dynamic Offer Set Extension}}
In our basic problem formulation, we assumed that all buying and selling offers $\calB$ and $\calS$ are available at once, and we cleared the market in one take.
In practice, however, 
%
% both the prosumers and the DSO may continuously submit new offers as their predictions, physical state, and the market conditions change over time. This means that we need to re-solve intervals as new information becomes available since the optimal solution to the energy trading problem may change, and the optimal value of each $p_{s,b,t}$ may vary.
%
the market conditions and the physical state of the DSO and prosumers may change over time, making it advantageous to submit new offers. \iffuture{Updating or cancelling offers could also be beneficial, however we do not provide that functionality here, but we describe how it could be accomplished in Section \ref{sec:future}. }\fi As new offers are posted we need to recompute the solution.
While new offers can increase the amount of energy traded, the \emph{trade values} $p_{s,b,t}$ and $\pi_{s,b,t}$ need to be \emph{finalized} at some point in time.
At the very latest, values for interval $t$ need to be finalized by the end of interval $t - 1$; otherwise, participants would have no chance of actually delivering the trade. 

Here, we extend the energy trading problem to accommodate a time-varying offer set (where offers can be unmatched, matched and pending, or matched and finalized), and a time constraint for finalizing trades. Our approach finalizes only trades that need to be finalized, which maximizes efficiency while providing safety.
%Consequently, the optimal matching changes over time as new offers arrive.
%A simple approach for accommodating an evolving set of offers would be to periodically find an optimal matching, finalize the trades, and start from scratch for the next set of offers. 
%In this section, we propose a superior approach, which improves trading by matching as many offers at the same time as possible, while finalizing as few of them as possible. 
%
% \ad{clarify that an offer can change if not matched and new offers can show up for a future time. Also an offer can be matched and then canceled by the system -- unless it has been finalized-- if a better match is found. you might want to clarify the states of an offer}
%
We assume that all trades for time interval $t'$ (i.e., all values~$p_{s,b,t'}$ and $\pi_{s,b,t'}$) must be finalized and the trading prosumers must be notified by the end of time interval $t' - T_{clear} - 1$ (see Fig. \ref{fig:timeline}), where $T_{clear}$ is a positive integer constant that is set by the DSO. 
In other words, if the current interval is $t$, then all intervals up to $t + T_{clear}$ have already been finalized.
Preventing ``last-minute'' changes can be crucial for safety and fairness since it allows both the DSO and the prosumers to prepare for delivering (or consuming) the right amount of power. In practice, the value of $T_{clear}$ must be chosen accounting for both physical constraints (e.g., how long it takes to turn on a generator) and communication delay (e.g., some participants might learn of a trade with delay due to network disruptions).

We let $\hat{p}_{s,b,t}$ and $\hat{\pi}_{s,b,t}$ denote the finalized trade values, and we let $\calB^{(t)}$ and $\calS^{(t)}$ denote the set of buying and selling offers that participants have submitted by the end of time interval $t$.
Then, the system takes the following steps at the end of each time interval $t$.
%\begin{itemize}
%\item 
First, find an optimal solution $(\vp^*, \vpi^*)$ to the extended energy trading problem:
\begin{equation}
\max_{(\vp, \vpi) \, \in \textit{ Feasible}(\calS^{(t)}\!,\, \calB^{(t)})} \sum_{s \in \calS^{(t)}} \sum_{b \in \calM(s)} \sum_{\tau \in I(s,b)} p_{s,b,\tau}
%\max_{\substack{(\vp, \vpi) \, \in \textit{ Feasible}(\calS^{(t)}\!,\, \calB^{(t)}) \\ \text{subject to} \\ \forall \tau \leq t - T_{clear}: ~ \hat{p}_{s,b,t} = p_{s,b,t} \wedge \hat{\pi}_{s,b,t} = \pi_{s,b,t}}}
\end{equation}
subject to
\begin{align}
\forall \, \tau \leq t_{f}: ~ & p_{s,b,\tau} = \hat{p}_{s,b,\tau} \\ 
% \forall \, \tau \leq t + T_{clear}\!: ~ & p_{s,b,\tau} = \hat{p}_{s,b,\tau} \\
& \pi_{s,b,\tau} = \hat{\pi}_{s,b,\tau} 
\end{align}
%\item 
Second, finalize trade values for time interval $t_{f}$ based on the optimal solution $(\vp^*, \vpi^*)$:
\begin{align}
% & \hat{p}_{s,b,t + T_{clear}} := p^*_{s,b,t + T_{clear}} \\ 
% & \hat{\pi}_{s,b,t + T_{clear}} := \pi^*_{s,b,t + T_{clear}} 
& \hat{p}_{s,b,t_{f}} := p^*_{s,b,t_{f}} \\ 
& \hat{\pi}_{s,b,t_{f}} := \pi^*_{s,b,t_{f}} 
\end{align}
%\end{itemize}

By taking the above steps at the end of each time interval, trades are always cleared based on as much information as possible (i.e., considering as many offers as possible)\footnote{This includes offers for intervals beyond the finalization interval. Effectively, matches for an interval beyond finalization can be changed if a better solution is found; however, finalized matches are permanent and never changed.} without violating any safety or timing constraints. Note that here $\textit{Feasible}(\calS, \calB)$ now also includes the safety constraints~\eqref{eq:EL}, \eqref{eq:constrInt}, and \eqref{eq:constrExt}.

\begin{figure}
    \centering
    \begin{tikzpicture}

    \node(s) at (0,0) {};
    \node(d) at (5,0) {};
    
    \node(tc) at (.25,-.3) {$t$};
    \node(tf) at (2.25,-.3) {$t_f$};
    
    \draw[-latex](s.center) -- (5.5,0) {};
    
    % draw vertical lines
    \foreach \x in {0,.5,...,5} 
    \draw (\x cm,3pt) -- (\x cm,-3pt);
    
    \draw[decorate,decoration={brace,amplitude=5pt}] (.5,0.3) -- (2,0.3)
    node[anchor=south,midway,above=4pt] {$T_{clear}$};
    
    \draw[decorate,decoration={brace,amplitude=5pt}] (2.5,0.3) -- (4,0.3)
    node[anchor=south,midway,above=4pt] {$T_{h}$};
    \draw[dashed,semithick,latex-latex] (3.5,0.25) -- (4.5,0.25);
    
    % \draw[decorate,decoration={brace,amplitude=5pt}] (3.5,-0.6) -- (2.5,-0.6)
    % node[anchor=north,midway,below=4pt] {$T_{predict}$};
    % \draw[dashed,thick,latex-latex] (3,-0.5) -- (4,-0.5);
    
\end{tikzpicture}
    \caption{Temporal parameters ($t$ is current interval, $t_f$ is the interval to be finalized).}
    \label{fig:timeline}
\end{figure}
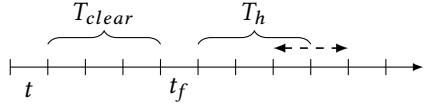

\subsection{Linear-Programming Solution:} \label{sec:linear-soln}

To find the optimal solution efficiently, we frame the energy trading problem  as a linear program. First, we create real-valued variables $p_{s,b,t}$ and $\pi_{s,b,t}$ for each $s \in S, b \in \calM(s), t \in I_s \cap I_b$.
Then, the following reformulation of the matching problem is a linear program:
\begin{equation}
\max_{\vp,\vpi}
\sum_{s \in \calS} \sum_{b \in \calM(s)} \sum_{t \in I(s,b)} p_{s,b,t} \label{eq:linProgObj}
\end{equation}
subject to the constraint Equations, which can all be expressed as linear inequalities~\eqref{eq:constrEnergy}, \eqref{eq:constrPrice}, \eqref{eq:EL}, \eqref{eq:constrInt}, \eqref{eq:constrExt}, and
\begin{equation}
\vp \geq \vect{0} \text{ and } \vpi \geq \vect{0} .
\end{equation}.

\subsubsection{Practical Considerations}
Even though Equation \eqref{optimalproblem} can be formulated as a linear program and be solved efficiently (i.e., in polynomial time), the number of variables $\{p_{s,b,t}\}$ may grow prohibitively high as the number of offers and time intervals that they span increases.
In practice, this may pose a significant challenge for solving the energy trading problem  for larger transactive microgrids.
A key observation that helps us tackle this challenge is that even though prosumers may post offers whose latest intervals are far in the future,
%(i.e., for an offer $s$, the latest interval may be $\max(I_s) \gg t_{c}$, where $t_c$ is the current interval)
the optimal solution for the finalized interval typically depends on only a few intervals ahead of the finalization deadline. 
Indeed, we have observed that considering intervals in the far future has little effect on the optimal solution for the interval that is to be finalized next (see Fig. \ref{fig:loahNrgMEM}).

% % \begin{figure}
% %     \centering
% \begin{wrapfigure}{r}{.5\textwidth}
%     \includegraphics[width=1.0\linewidth]{figs/2018-9-14widowSizeVsEnergyTraded.png}
%     \caption{Memory consumption and Energy traded during a single interval of the simulation (interval 80) for various values of $T_{lookahead}$.}
%     \label{fig:loahNrgMEM}
% \end{wrapfigure}
% %source: https://docs.google.com/spreadsheets/d/1EoQKCElVXEhQFYyZa_DSOTMs6ox_NBGuxqHz0NxtSew/edit?usp=sharing
% % \end{figure}
    
Consequently, for practical solvers, we introduce a planning horizon $T_{h}$ (see Fig. \ref{fig:timeline}) that limits the intervals that need to be considered for a solution: for any $\hat{t} > t_f + T_{h}$, we set $p_{s,b,\hat{t}} = 0$, where~$t_{f}$ is the earliest interval that has not been finalized. By ``pruning'' the set of free variables, we can significantly improve the performance of the solver with negligible effect on solution quality (see Fig. \ref{fig:loahNrgMEM}).
This results in the following ``pruned'' objective function:
%We let $\hat I(s, b)$ be the set of intervals before $t_f + T_h$ in the intersection of offers $s$ and $b$ (i.e., $\hat I(s, b) = I_s \cap I_b \cap \{\tau; \tau \leq t_{f} + T_{h}\}$, this modifies Equation \eqref{optimalproblem} to:
\begin{align}
\label{ext_prob1}
\max_{(\vp,\vpi) \, \in \textit{ Feasible}(\calS, \calB)} ~
\sum_{s \in \calS} \sum_{b \in \calM(s)} \sum_{\tau \in I_s \cap I_b \cap \{\tau; \tau \leq t_{f} + T_{h}\}} p_{s,b,\tau}
\end{align}

% There are 3 questions that need to be addressed when implementing this trading approach, they are:
% \begin{enumerate}
%     \item What is the value of $T_{clear}$ ?
%     \item How should groups be formed? 
%     \item What value of $T_{h}$ should a solver use?
% \end{enumerate}
% In Section \ref{sec:analysis} we discuss question 2. %In Section \ref{sec:experimental} we share what values we chose for $T_{clear}$ and $T_{h}$ and why.

% The second issue is that Equation \eqref{optimalproblem} formulates the problem considering a single ``snapshot'' of all offers across all time intervals. However, in practice, prosumers may submit new offers at any time, resulting in continuously evolving sets of offers. Consequently, optimal solutions to Equation \eqref{optimalproblem} may have to be found repeatedly as new offers are submitted, resulting in a series of evolving solutions. This presents a problem since prosumers need to know in advance what the ``final'' solution for a certain time interval is in order to be able to actually schedule energy production or consumption for that interval. Further, preventing ``last-minute'' changes can be crucial for safety and stability since it allows the DSO to prepare for satisfying energy demand that cannot be met locally.

\section{\platform Components}
\label{sec:components}

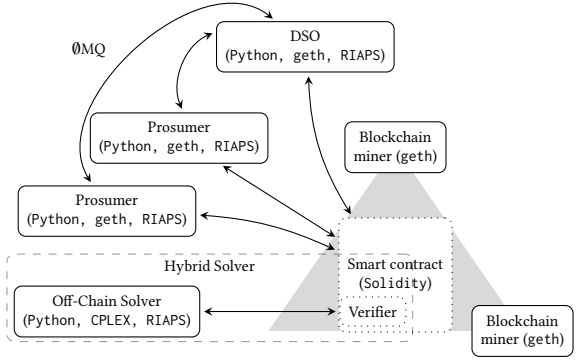
\begin{figure}[t]
\centering
%\resizebox{\columnwidth}{%
\begin{tikzpicture}[x=1.2cm, y=1.2cm, font=\tiny,
  Component/.style={fill=white, draw, align=center, rounded corners=0.1cm},
  Connection/.style={<->, >=stealth, shorten <=0.05cm, shorten >=0.05cm}]
 
\fill [fill=black!15] (90:1.5) -- (200:1.5) -- (340:1.5) -- (90:1.5);

\foreach \pos/\name in {1.6/pros3, 0.8/pros2} { %, 0.0/pros1} {
  \node [Component] (\name) at (\pos - 4, \pos) {Prosumer\\(\texttt{Python, geth, RIAPS})};
}

\node [Component] (dso) at (-1, 2.6) {DSO\\(\texttt{Python, geth, RIAPS})};

\node [Component] (ocsolver) at (-3.2, -.3) {Off-Chain Solver\\(\texttt{Python, CPLEX, RIAPS})};

\foreach \pos in {90, 340} {
  \node [Component] at (\pos:1.5) {Blockchain\\ miner (\texttt{geth})};
}

\node [Component, dotted, minimum height=1.5cm] (contract) at (0, .1) {Smart contract\\(\texttt{Solidity})};

\node[dashed, opacity=0.4,text opacity=1,Component, fill=none,minimum width=5.4cm,minimum height=1.15cm, label={[label distance=-.4cm]90:{Hybrid Solver}}](hybrid) at (-2.06,-.15){};

\node[Component,dotted](verifier) at (-.25,-.3) {Verifier};

%\draw [Connection, bend left=45] (solver) to (dso);

%\draw [Connection, bend left=65] (pros1) to node [midway, left, shift={(-0.25,0)}] {$\emptyset$MQ} (dso);
\draw [Connection, bend left=80] (pros2) to node [midway, left, shift={(-0.25,0)}] {$\emptyset$MQ} (dso);
\draw [Connection, bend left=45] (pros3) to (dso);

\draw [Connection] (ocsolver) to node [above] {} (verifier);
\draw [Connection, bend right=15] (dso) to (contract);
%\draw [Connection, bend right=0] (pros1) to node [midway, below left] {Ethereum} (contract);
\draw [Connection, bend right=-10] (pros2) to (contract);
\draw [Connection, bend right=0] (pros3) to (contract);
\end{tikzpicture}%
%}
\caption{Components of the energy trading system. In our reference implementation, we use Ethereum as the decentralized computation platform for smart contracts, and the other components interact with the blockchain network using the \texttt{geth} Ethereum client. The smart contract is implemented in Solidity, a high-level language for Ethereum, and it is executed by a private network of \texttt{geth} mining nodes. The off-chain solver uses CPLEX. %\textcolor{red}{is the verifier a function in smart contract. I dont think we have explained it in text}\textcolor{blue}{its section \ref{sec:contract-verify}}
}
\label{fig:components}
\end{figure}
In this section, we introduce the components of \platform (see Fig.~\ref{fig:components}), which implement the functionality described in the previous section. The components are built on top of RIAPS~\cite{eisele2017riaps,Volgyesi2017Time}.
For brevity, we do not elaborate on the role of RIAPS in this paper, and refer interested readers to~\cite{transax}. % The details of time synchronization, a key feature in the design is presented in \cite{Volgyesi2017Time}.}

%

% In this section we describe the technology underlying the components that constitute the platform depicted in Fig. \ref{fig:components}, followed by a discussion of each of the components. 

% \subsection{Communication Middleware}

% The components are built on top of RIAPS (see Section \ref{sec:riaps}) enabling them to communicate and synchronize the actual energy transfer which is critical in power systems and is presented in greater detail in \cite{Volgyesi2017Time}. 

% In the component discussions below we have included the component description code that was written using the RIAPS domain-specific language. We briefly describe the relevant keywords that appear. In the DSO we see the \textit{rep} keyword which initiates the component with a reply port. \textit{contractAddr} is the name of the port, in the prosumer snippet we again see a port called \textit{contractAddr}, which has the keyword \textit{req} this is the corresponding request port. On deployment RIAPS configures the communication between these two components over this port. The other keyword we are interested in is \textit{timer} which instantiates the component with a timer that triggers a function call. In the DSO the timer calls the \textit{finalizer} function every 2 minutes.
% \Aron{How is this relevant to the contributions of this paper? We might as well be describing what Python libraries we import.}\ad{Typically, a systems paper focuses on the details of ports and rates. Anyways given the space limitations  these can be commented out}

\subsection{DSO}

\iflong{
\begin{lstlisting}[language=riaps,numbers=none]
    component DSO(logfile = "DSO") {
    req contractAddr :: (Query_contract_address,Contract_address) ;
    timer finalizer 120000 ms;
    } 
\end{lstlisting}
}
\fi

% THIS IS IN THE ``ENERGY APPROACH'' SECTION
% We assume the existence of a distribution system operator (DSO), which also participates in the market. The DSO may use the market to incentivize timed energy production within the microgrid to aid in grid stabilization and in the promotion of related ancillary services \cite{7462854} through updates to the price policy. In addition, the DSO supplies residual demand not met through the local market. 

The DSO is a trusted entity that manages the microgrid, handling financial operations like sending monthly bills, and functional operations like registering new smart meters. %, and setting the energy constraints of new customers.

Monthly bills are handled in conjunction with the smart meter. 
The DSO meets the prosumers' residual demand and supply, i.e., consumption and production that they did not trade in advance due to inaccurate predictions or lack of trade partners.
For each interval, the DSO sets the prices $\pi^{S}_{t}$ and $\pi^{B}_{t}$ for this residual energy consumption and production, respectively, which will be used by the smart meter for billing (see Section~\ref{sec:billing}). 

Registration means adding a new smart meter to the platform when it is installed for a prosumer. Registration includes assigning the smart meter to the prosumer's group and %, which is the prosumer's feeder by default.
%Registration also includes 
setting the prosumer's limits $\textit{EPL}$ and $\textit{ECL}$ according to the amount of energy the hardware connected to the smart meter can feasibly produce and consume respectively. The DSO can update these limits as required due to changes in the hardware monitored by the smart meter. 
% Finalizing trades occurs every interval which causes any further solutions for that interval to be rejected. 

\subsection{Prosumers}
\label{sec:prosumer}

\iflong
\begin{lstlisting}[language=riaps,numbers=none]
        component Trader(ID=101, logfile="") {
        req contractAddr : (Query_contract_address,Contract_address);
        timer poller 1000 ms;
        timer post 5000 ms;
        }
\end{lstlisting}
\fi

% \iAron{These assumptions should have been introduced much earlier!}\iScott{Leave here for now}
% Prosumers are located in a specific feeder, and they are assigned to that feeder's group, which may consist of multiple feeders.
Prosumers are able to withdraw assets for each interval from the platform up to the limits set by the DSO, and trade those assets with other prosumers.
%It is possible that if all the energy represented by the assets allocated to the prosumers were utilized it would damage the feeder, but the group safety constraints prevent this. 
To protect their privacy, they may also create new anonymous accounts and use them for trading. To maintain the anonymity of these accounts, prosumers always transfer assets to them through a \textit{anonymizing mixer}. The mixer ensures that accounts cannot be linked to the prosumer that owns them. The mixer is implemented as a protocol that happens between prosusumers, not through the DSO or other central entity. We discuss mixing and how it fits in the workflow in Section \ref{sec:mixing}.

\subsection{Smart Meter}
\label{sec:smartmeter}
The role of the smart meter is to measure the prosumer's energy production and consumption, to monitor the assets allocated to the prosumer, and to provide privacy-preserving billing. 

The smart meter checks that all assets withdrawn by a prosumer are accounted for. It also measures how much energy the prosumer transferred, ensuring that it does not exceed the $ECL$ or $EPL$ limits, and how much was not part of a trade. The amount that was not accounted for in trades is used to compute how much the prosumer owes the DSO for the interval. Each billing cycle (e.g., monthly) the smart meter sends that information to the DSO so it can bill the prosumer. This way the DSO has no fine-grained information on the energy profile of the consumer, it only knows the amount it needs to be paid for the energy it provided during that cycle. We discuss the workflow of the smart meter in Section~\ref{sec:billing}.

% When a prosumer withdraws assets, the smart meter records the amount withdrawn. When an interval is finalized, the prosumer transfers the assets it acquired through trading and those that were not traded away to an anonymous account that belongs to the smart meter. This way, only the smart meter knows the final asset balance of the prosumer. 

% \iAron{Smart meter needs to keep track of only the production assets: the net energy bought by a prosumer is simply the difference between total production assets at the end minus total production assets withdrawn.}

% Then during an interval the smart meter measures the actual energy transferred and any difference between the actual amount and the assets acquired through trades is added to the prosumers bill. For each billing cycle (e.g., every month) the smart meter then sends that information to the DSO so it can bill the prosumer. This way the DSO has no fine-grained information on the energy profile of the prosumer, only the cumulative energy trades that occured with the DSO itself. Note that even though we do not discuss penalizing prosumers who do not conform to the solution, it is straightforward to do this however it should be considered only if the discrepancy is extraordinary.
% \iScott{Does this seem too much like what is in the workflow? }
% \iAron{Remove details, reference Section~\ref{sec:billing}.}
    
\subsection{Smart Contract}
\label{sec:contract}

% \iAron{We need to emphasize the smart-contract properties that we use in this paper: (1) trustworthy execution, (2) general-purpose computation (quasi-Turing complete), (3) financial settlements, and (4) could be implemented on various platforms, not just distributed ledgers (we use a distributed ledger based smart contract for the sake of decentralization).}
% \iAron{We need to say all that at some point, and this is the first mention of smart contracts, so let's do it here.}

% \iAron{do we explain financial transfers anywhere in the paper?}
% \iScott{here, and we mention the smart contract transfers funds in Section \ref{sec:billing}. I think that is all}

Smart-contracts are programs that provide trustworthy, general-purpose computation. Since they are trusted, they are suitable for financial settlements and have been implemented on various platforms including distributed ledgers. Contracts that utilize distributed ledgers are beneficial because they enable the decentralized enforcement of the contract rather than relying on a single trusted entity. We use a blockchain-based smart contract, since it provides a distributed ledger, to enable sensitive market features like asset trading and financial transactions.

% the smart-contract properties that we use in this paper: (1) trustworthy execution, (2) general-purpose computation (quasi-Turing complete), (3) financial settlements, and (4) could be implemented on various platforms, not just distributed ledgers (we use a distributed ledger based smart contract for the sake of decentralization).

\iflong
The functions and their respective events can be seen in Table \ref{tab:Events}.
\fi
The distributed ledger provides immutable storage service for offers, solutions, safety constraints, and a notification log of events. Prosumers and solvers check for these events and perform actions based on them.
Offers are posted to the ledger using the smart contract functions. Valid offers must simply contain the format specified in Section \ref{sec:approach} and the account posting the offer must have assets to cover what is offered. If an offer is valid, it is accepted and made available to other services through the smart contract.

\iflong
\begin{table}[!htbp]
    \centering
    
    \caption{Smart Contract Functions and Events.}
    \begin{tabulary}{\textwidth}{|L|L|L|}
    \hline Function & Event & Caller \\
    % \hline setup & NA & On deployment\\
    \hline registerProsumer & ProsumerRegistered & Prosumer\\
    \hline withdrawAssets & AssetsWithdrawn & Prosumer \\
    \hline postBuyingOffer & BuyingOfferPosted & Prosumer\\
    \hline postSellingOffer & SellingOfferPosted & Prosumer\\
    \hline postSolution & SolutionPosted & Solver\\
    \hline finalize & Finalized & DSO \\
    \hline
    
    \end{tabulary}

    \label{tab:Events}
\end{table}
\fi

\subsection{Hybrid Solver}
\label{sec:hybridSolver}

Although solving linear programs is not computationally hard, it can be challenging with a large number of variables and constraints in resource-constrained computing environments. Since computation is relatively expensive on blockchain-based distributed platforms\footnote{Further, Solidity, the preferred high-level language for Ethereum, currently lacks built-in support for certain features that would facilitate the implementation of a linear programming solver, such as floating-point arithmetics~\cite{Wood2014ethereum}.}, solving even  the ``pruned'' energy trading problem from Equation \eqref{ext_prob1} might be infeasible using a block\-chain-based smart contract.
In light of this, we propose a \emph{hybrid implementation approach}, which combines the trustworthiness of blockchain-based smart contracts with the efficiency of more traditional computational platforms.

The key idea of our hybrid approach is to (1) use high-performance computers to solve the computationally expensive linear program \emph{off-blockchain} and then (2) use a smart contract to record the solution \emph{on the blockchain}.
To implement this hybrid approach securely and reliably, we must account for the following issues.
% \Aron{If we have space, we can add some left margin for the lists (but let's leave this for the very end).}
\begin{itemize}[topsep=0pt, leftmargin=1em]
\item Computation that is performed off-blockchain does not satisfy the auditability and security requirements that smart contracts do. Thus, the results of any off-blockchain computation must be verified in some way by the smart contract before recording them on the blockchain.
\item Due to network disruptions and other errors (including deliberate denial-of-service attacks), the off-blockchain solver might fail to provide the smart contract with a solution on time (i.e., before trades are supposed to be finalized). Thus, the smart contract must be able to proceed without an up-to-date solution.
\item For the sake of reliability, the smart contract should accept solutions from multiple off-blockchain sources; however, these sources might provide different solutions.
Thus, the smart contract must be able to choose from multiple solutions (some of which may come from a compromised computer).
\end{itemize}

%\Abhishek{The time synchronization discussion can go here.}
\subsubsection{Smart Contract based Solution Verification}
\label{sec:contract-verify}

Apart from storing data (Section \ref{sec:contract}), our smart contract is also designed to (1) verify whether a solution $(\vp, \vpi)$ is feasible and (2) compute the value of the objective function for a feasible solution.
Compared to finding an optimal solution, these operations are computationally inexpensive, and they can easily be performed on a blockchain-based decentralized platform.
Using these capabilities, the smart contract provides the following functionality:
\begin{itemize}[topsep=0pt, leftmargin=1em]
\item Solutions may be submitted to the contract at any time. 
The contract verifies the feasibility of each submitted solution, and if the solution is feasible, then it computes the value of the objective function.
The contract always keeps track of the best feasible solution submitted so far, which we call the \emph{candidate solution}.
\item At the end of each time interval $t$, the contract finalizes the trade values for interval 
$t_{f} = t + T_{clear} + 1$ based on the candidate solution.\footnote{If no solution has been submitted to the contract so far, which might be the case right after the trading system has been launched, $\vp = \vect{0}$ may be used as a candidate solution.}
\end{itemize}

% This simple functionality achieves a high level of security and reliability.
% Firstly, it is clear that an adversary cannot force the contract to finalize trades based on an unsafe (i.e., infeasible) solution since such a solution would be rejected.
% Similarly, an adversary cannot force the contract to choose an inferior solution instead of a superior one.
% In sum, the only action available to the adversary is proposing a superior feasible solution, which would actually improve energy trading in the microgrid.

% The contract is also reliable and can tolerate temporary disruptions in the solver or the communication network.
% Notice that any solution $(\vp, \vpi)$ that is feasible for sets $\calS$ and $\calB$ is also feasible for supersets $\calS' \supseteq \calS$ and $\calB' \supseteq \calB$.
% As the sets of offers can only grow over time, the contract can use a candidate solution submitted during time interval $t$ to finalize trades in any subsequent time interval $\tau > t$.
% In fact, without receiving new solutions, the difference between the amount of finalized trades and the optimum will increase only gradually:
% since the earlier candidate solution can specify trades for any future time interval,
% the difference is only due to the offers that have been posted since the solution was found and submitted.

%In our hybrid approach, we implement a relatively simple algorithm in our smart contract, which can (1) check if a solution $(\vp, \vpi)$ is feasible and (2) evaluate the objective function for a given solution (see Equation~\eqref{eq:linProgObj}).

\subsubsection{Off-blockchain Matching Solver}

We complement the smart contract with an efficient linear programming solver, specifically CPLEX\footnote{https://www.ibm.com/analytics/cplex-optimizer}, which can be run off-blockchain, on any capable computer (or multiple computers for reliability).
The solver is run periodically to find a solution to the energy trading problem based on the latest set of offers posted. 
Once a solution is found by the matching solver, it is submitted to the smart contract in a blockchain transaction.
Note that if new offers have been posted since the solver started working on the solution, the contract will still consider the solution to be feasible. 
This is because any feasible solution for sets $\calS$ and $\calB$ also being feasible for supersets $\calS' \supseteq \calS$ and $\calB' \supseteq \calB$.
%The smart contract then verifies whether the solution is feasible and whether it is better than any feasible solution submitted before,  and if both conditions hold, accepts the solution as the best matching (found so far).

From the perspective of the solver, being able to submit multiple solutions to the contract for the same problem has many advantages. %that multiple solutions may be submitted to the smart contract, which will always choose the best feasible one.
%This is advantageous because
For example, it allows the linear programming solver to be run as an anytime algorithm.
Further, we can allow multiple---potentially untrusted---entities to try to solve the problem and submit solutions, since the smart contract will always choose the best feasible one.
This is especially important in microgrids where a trusted third party is not guaranteed to always be present.
In such settings, prosumers can be allowed to volunteer and provide solutions to the energy trading problem. \footnote{Of course, each prosumer will try to submit a solution that favors the prosumer. However, the submitted solution still needs to be superior with respect to the optimization objective, which roughly corresponds to social utility. Hence, each prosumer is incentivized to improve social utility by submitting superior solutions that favor the prosumer. We leave the analysis of these incentives for future work.}

Thereby, we enable finding solutions in an efficient and very flexible manner, while reaping the benefits of smart contracts, such as auditability and trustworthiness.

\section{\platform Protocol}
\label{sec:protocol}

\begin{figure}
    \centering
    \scalebox{1.2}{
    \begin{tikzpicture}[x=.1cm, y=.1cm, font=\tiny,
        Component/.style={fill=white, draw, align=center, rounded corners=0.1cm},
        Connection/.style={<->, >=stealth, shorten <=0.05cm, shorten >=0.05cm},
        call/.style={->, >=stealth, shorten <=0.05cm, shorten >=0.05cm},
        event/.style={<-, >=stealth, shorten <=0.05cm, shorten >=0.05cm}]
         
        \fill [fill=black!15] (90:25) -- (200:25) -- (340:25) -- (90:25);
        
        \node [Component, dotted, minimum height=2.75cm] (contract) at (0, 5) {Smart contract\\(\texttt{Solidity})};
        
        \node[Component, minimum height=1.6cm] (p1) at (-40,9){Prosumer 1};
        \node[Component, minimum height=10] (p2) at (-40,25){Prosumer 2};

        \node [Component] (dso) at (0, 30) {DSO};
        
        \node[dashed, fill opacity=0.4,text opacity=1,Component,fill=none,minimum width=5.2cm,minimum height=1.1cm, label={[label distance=-.4cm]90:{Hybrid Solver}}](hybrid) at (-23,-6){};
        
        \node [Component] (ocsolver) at (-40, -5) {Off-Chain Solver};
        \node[Component,dotted](verifier) at (-3,-5) {Verifier};

        \draw[call]([shift={(-2,0)}]dso.south) to node [xshift=-1cm] {\circled{1} registerSmartMeter}([shift={(-2,0)}]contract.north);
        \draw[call]([shift={(2,0)}]dso.south) to node [xshift=0.5cm] {\circled{9} finalize}([shift={(2,0)}]contract.north);
        % \draw[call](dso.east) -| ([xshift=0.9cm]contract.north);
        
        \draw[Connection]([shift={(0,0)}]p2.south) to node [xshift=0.3cm] {\circled{4}  mix}([shift={(0,0)}]p1.north);
        
        \draw[call]([shift={(0,16-8)}]p1.east) to node [yshift=.1cm] {\circled{2} registerProsumer}([shift={(0,16-4)}]contract.west);
        \draw[call]([shift={(0,13-8)}]p1.east) to node [yshift=.1cm] {\circled{3} withdrawAssets} ([shift={(0,13-4)}]contract.west);
        % \draw[call]([shift={(0,11-8)}]p1.east) to node [yshift=.1cm] {\circled{4} createAnonAddr} ([shift={(0,11-4)}]contract.west);
        \draw[call]([shift={(0,10-8)}]p1.east) to node [yshift=.1cm] {\circled{5} postOffer} ([shift={(,10-4)}]contract.west);
        \draw[event]([shift={(0,7-8)}]p1.east) to node [yshift=.1cm] {\circled{8} SolutionPosted} ([shift={(0,7-4)}]contract.west);
        \draw[event]([shift={(0,3.5-8)}]p1.east) to node [yshift=.12cm] {\circled{10} Finalized} ([shift={(0,3.5-4)}]contract.west);
        \draw[call]([shift={(0,-8)}]p1.east) to node [yshift=.11cm] {\circled{11} deposit} ([shift={(0,-4)}]contract.west);
        
        \draw[->,dashed, align=center]([shift={(0,0)}]p1.west) -|  node [xshift=-.6cm, yshift=0.7cm] {\circled{12} Energy \\ Transfer} (-54,20) |- ([shift={(0,0)}]p2.west);
        
        \node [Component,minimum width=1.5cm] (sm1) at (-61, 10) {Smart Meter 1};
        \node [Component] (sm2) at (-60, 25) {Smart Meter 2};

        % \draw[call]([shift={(0,-6)}]p1.west) -| node [xshift=.5cm, yshift=.1cm] {13) request} ([shift={(0,0)}]sm1.south);

        \draw[call]([shift={(4,0)}]contract.south) |-  node [xshift=-2cm, yshift=-.10cm] {\circled{6} OfferPosted} (-20,-13) -| ([shift={(3,0)}]ocsolver.south);
        \draw[call]([shift={(6,0)}]contract.south) |-  node [xshift=-2cm, yshift=-.14cm] {\circled{10} Finalized} (-20,-16) -| ([shift={(-3,0)}]ocsolver.south);
        
        \draw [call] (ocsolver) to node [yshift=-.1cm] {\circled{7} postSolution} (verifier);

        \draw[call]([shift={(-7,0)}]sm1.north)|- node [xshift=2cm,yshift=.12cm]{\circled{13} bill} (dso.west);

    \end{tikzpicture}%
    }
    \caption{Example workflow of \platform. Nodes represent entities in the platform, and edges represent interactions, such as smart-contract function calls. In this example, prosumer 1 is selling energy to prosumer 2 and the dashed line represents the energy transfer.}
    \label{fig:workflow}
    
\end{figure}
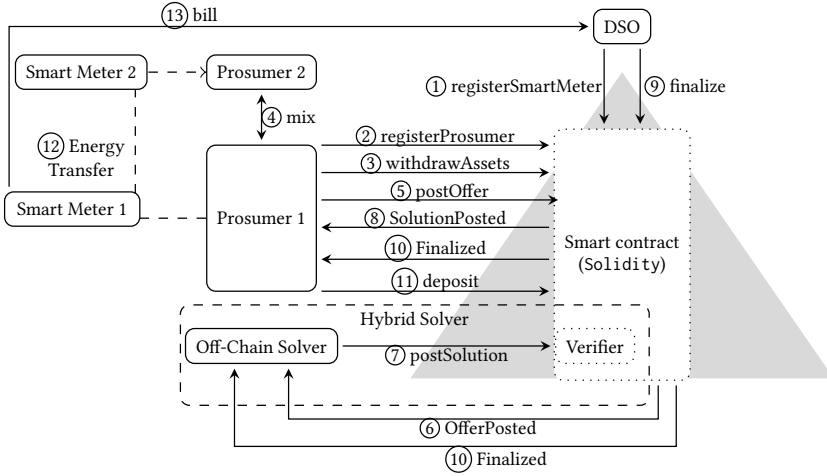

In this section, we  describe how the components introduced in the previous section interact via the \platform protocol, which is depicted in Fig.~\ref{fig:workflow}

\subsection{Registration}

When a new customer is added to the grid, a smart meter is installed. The DSO registers the smart meter by calling \circled{1}\footnote{The circled numbers correspond to the numbered edges in Fig.~\ref{fig:workflow} } \textit{registerSmartMeter} on the \platform smart contract. This call sets the asset allocation limits for that customer and records which feeder it is located on in the grid. The customer then registers as a prosumer with \platform by calling \circled{2} \textit{registerProsumer}. 

The registration information requires each prosumer to specify a smart meter, and to provide a DSO certified public address that corresponds to the specified smart meter for the DSO to use when allocating assets. Since the smart meter is associated with a specific feeder, the smart contract adds the prosumer to the group associated with that feeder. This is required to ensure that feeder-level safety constraints can be correctly applied. The registrations can happen asynchronously, allowing new prosumers to  join at any time, even long after trading has commenced. The registration process occurs only once for each smart meter and prosumer. Once registered, a prosumer may participate in the following trading protocol repeatedly.

\subsection{Mixing}
\label{sec:mixing}

Once a prosumer has registered, it may withdraw assets into their public address for future intervals by calling \circled{3} \textit{withdrawAssets}. After withdrawing assets, a prosumer could make offers using \textit{postOffer}. However, if it made offers using its public account, then the trades could be traced back to the prosumer, thereby violating the privacy requirements. Instead, the prosumer creates anonymous addresses and transfers the assets from its public address to the anonymous addresses via \circled{4} mixing assets with other prosumers. They can do this within their assigned groups by executing a decentralized mixing protocol such as the one described in \cite{ruffing2014coinshuffle}. Within the group, each prosumer supplies anonymous addresses and the amount to be deposited in each address. They also supply a public address containing the assets that will be transferred to the anonymous addresses. Then, the prosumers execute the protocol, transferring the assets anonymously, without associating them with the public addresses.  If the prosumer transferred assets to a new ``anonymous'' account directly without using mixing, then the new account would not truly be anonymous as there would be no difficulty in tying it back to the prosumer. 

\subsection{Trading}

Now the prosumers can post anonymous offers using their anonymous accounts by calling function \circled{5} \textit{postOffer}. The smart contract checks that the account posting the offer has assets that cover the amount and intervals specified in the offer. If not then the offer is rejected. If the offer is valid the smart contract then emits event \circled{6} \textit{OfferPosted}, notifying the off-chain matching solvers. The matching solvers may wait for many prosumers to post many offers, but eventually, it pairs buying and selling offers and posts the solutions by calling function \circled{7} \textit{postSolution}. The smart contract checks the solution to make sure that it is feasible according to the feasibility requirements described in Section \ref{sec:approach}, including checking that the trades do not exceed the group capacity constraint. If the solution is valid, then smart contract saves it and emits event \circled{8} \textit{SolutionPosted}, notifying the prosumers of the current candidate solution. Additional solutions may be submitted by any solver, and if those solutions are valid and superior, \textit{i.e.}, they trade more energy, then the smart contract will update the candidate solution. 

\subsection{Billing}
\label{sec:billing}
As an interval comes to a close, the DSO calls\footnote{Note that by default the DSO calls the \textit{finalize} function to increment the current interval, but since this function is time guarded, any other entity can call it, which provides additional resilience.} function \circled{9} \textit{finalize} which means that offers for interval $t_f$ are no longer accepted and the smart contract transfers funds from the consuming offer's account to the producing offer's account. It also exchanges the $EPA$ assets of the seller for the $ECA$ assets of the buyer and vise versa for each of the matched offers. The call also emits the \circled{10} \textit{Finalized} event, notifying the solvers to update their solving interval, and the prosumers that the trades for interval $t_f$ have been finalized. If a prosumer posts offers with many anonymous accounts, it will have to aggregate all the corresponding trades to determine how much power it is expected to produce/consume during that interval when it arrives. Once the prosumers are notified of the trades, they call function \circled{11} \textit{deposit} to transfer all assets for the finalized interval from the prosumers anonymous accounts to an anonymous account owned by their smart meter. 

The smart meter checks that the total amount of assets deposited matches the amount withdrawn for the finalized interval. This ensures that there are no trades that have not been accounted for. It also compares the total of all production assets that were deposited against the production originally withdrawn to compute the net energy sold ($\Delta EPA = EPA_{\textnormal{\textit{deposit}}} - EPA_{\textnormal{\textit{withdraw}}}$). Then when the interval $t$ arrives and the power is \circled{12} transferred, the smart meter measures energy production/consumption. Let $E_u^t$ be net energy production (negative value represent net consumption) of prosumer $u$ at time interval $t$. The smart meter compares the net energy production (consumption) against the $EPL$ ($ECL$) of the prosumer to make sure it does not violate the safety constraint. It then computes the difference between the net energy sold and the net energy production to get the residual production (again, negative values are residual consumption). The residual production or consumption is multiplied by the DSO sell or buy prices, respectively, to compute what the prosumer owes the DSO each interval. Every \circled{13} billing cycle the smart meter sums the cost of the residuals and sends that to the DSO so the DSO can send the monthly bill. The bill $B_u^t$ of prosumer $u$ for timeslot $t$, which will be paid by the prosumer to the DSO, is
%
% \iAron{
% (1) smart meter measures measures energy production/consumption; let $E_u^t$ be net energy production (negative value represent net consumption); 
% (2) sum total production assets (deposited smart meter); compare sum total production assets to withdrawn production, to get net energy sold; 
%
% (3) compute difference net energy sold and net energy production, to get residual production (again, negative value is residual consumption; multiply residual production or consumption with DSO sell or buy prices, resp.}
%
\begin{align}
\label{eq:bill}
        B_u^t = &  \begin{cases} 
                \left(E_u^t + \Delta EPA \right) \cdot \pi^{S}_{t} & \text{ if } E_u^t + \Delta EPA < 0 \\
                  \left(E_u^t + \Delta EPA\right) \cdot \pi^{B}_{t} & \text{ otherwise,} \end{cases}
\end{align}
where $\pi^{S}_{t}$ is how much the DSO pays to purchase power and $\pi^{B}_{t}$ is how much the DSO charges for power. The prices could be functions of $E_u^t + \Delta EPA$, to charge more as the deviation from the predicted energy requirements increase. The price schedule is set for each timeslot $t$ by the DSO. 
% \Aron{I think that it should be $E_u^t + \Delta EPA$ since that's the deviation from forward trading (even if $\Delta EPA \neq$, deviation can be zero if $\Delta EPA = E_u^t$).}

% \iTaha{In the billing section, we only talked about the bills for excess amount of energy which is sold or bought by the DSO. How does money transfer between prosumers work? Is it through transferring Ether or another digital cryptocurrency? Or the DSO bills for this trades too?}
% \iAron{Yes, it would be good to discuss this in a sentence or two.}
% \iScott{This is addressed in the first sentence. The smart contract transfers funds from consuming to producing account when finalize is called.}
% $\pi^{S}_{t}$ and $\pi^{B}_{t}$

\section{Discussion and Analysis}
\label{sec:analysis}

In this section, we first describe how the \platform design ensures the security, resilience and safety of the system. Then, we provide a discussion on the inherent trade-offs between efficiency, and privacy.

%describe how \platform satisfies the requirements established in Section \ref{sec:intro}. 

%\begin{new}

% \subsection{Security, Safety, and Resilience }
\subsection{Requirement Evaluation}

    \subsubsection{Security and Safety}
    The underlying blockchain platform provides basic security features, so we are not concerned with the operations occurring on the blockchain. We are concerned with the secure and reliable operation of the solver. Similarly, the basic safety of the system is handled by the constrains described in Section \ref{sec:extend_safety}. The safety constraints are applied correctly and reliably by the same contract. An adversary cannot force the contract to finalize trades based on an unsafe (i.e., infeasible) solution since such a solution would be rejected. Similarly, an adversary cannot force the contract to choose an inferior solution instead of a superior one. In sum, the only action available to the adversary is proposing a superior feasible solution, which would actually improve energy trading in the microgrid.
   
   % This simple functionality of the smart contract presented in Section \ref{sec:contract-verify} achieves a high level of security and reliability.    Firstly, it is clear that an adversary cannot force the contract to finalize trades based on an unsafe (i.e., infeasible) solution since such a solution would be rejected. Similarly, an adversary cannot force the contract to choose an inferior solution instead of a superior one. In sum, the only action available to the adversary is proposing a superior feasible solution, which would actually improve energy trading in the microgrid.
    \subsubsection{Resilience}
    Now we show that our contract is reliable and can tolerate temporary disruptions in the DSO, solvers, or the communication network. First, since the \textit{finalize} contract function is time guarded any entity can call it, and the system can progress without a DSO which is only required for registering new prosumers and their smart metters. Second, notice that any solution $(\vp, \vpi)$ that is feasible for sets $\calS$ and $\calB$ is also feasible for supersets $\calS' \supseteq \calS$ and $\calB' \supseteq \calB$. As the sets of offers can only grow over time, the contract can use a candidate solution submitted during time interval $t$ to finalize trades in any subsequent time interval $\tau > t$. In fact, without receiving new solutions, the difference between the amount of finalized trades and the optimum will increase only gradually: since the earlier candidate solution can specify trades for any future time interval, the difference is only due to the offers that have been posted since the solution was found and submitted. Thus, the system can continue making trades using older valid solutions
    
  \subsubsection{Trading Efficiency}
  
    The trading platform we have presented is able to support efficient trading through temporal flexibility. We show this through Example~\ref{example:1}. As a reminder, this is due to prosumers being able to specify their production/consumption capacities and preferences (i.e., reservation prices) via offers and the linear-program finding an optimal matching. In Section \ref{sec:results}, we show using simulation  that energy trading reduces the load on the power grid.
    
    \begin{example}
    \label{example:1}
    Consider a community with two prosumers ($P_1$, $P_2$) and one consumer ($C_1$) on a single feeder. We divide each day into 15 minute intervals. Let us assume that $P_1$ has the ability to transfer $10$ kW into the feeder during interval 48, which translates to 12:00pm--12:15pm. Assume similarly that $P_2$ can also provide  $30$ kW to the feeder in interval 48, but it has battery storage. Since $P_2$ has battery---unlike $P_1$, who must either transfer the energy or waste it---$P_2$ can delay the transfer until a future interval, e.g., interval 49.  Now suppose that $C_1$ needs to consume $30$ kW in interval 48 and $10$ kW in interval 49. A possible solution would be to provide all $30$ kW to $C_1$ from $P_2$ in interval 48. However, that will lead to the waste of energy provided by $P_1$. Thus, a better solution will be to consume $10$ kW from $P_1$ in interval 48 and $20$ kW from $P_2$ in interval 48. Then, transfer $10$ kW from $P_2$ in interval 49, which is more efficient than the first matching as it allows more energy (summed across the intervals) to be transferred. Thus, we see that permitting temporal flexibility can significantly increase trading volume, though it does increase the size of the optimization problem, increasing computational complexity.  
    \end{example}

% \iAron{Mention that any component/entity can be offline for extended period of time, even the DSO (which is needed only for registering prosumers and their smart meters).}
\subsubsection{Privacy}

The platform provides pseudo-anonymity as the individual offers cannot be tied back to the prosumer who posted them since the offer is only affiliated with an anonymous address and contains only the energy amount and reservation price. Additionally the DSO does not know the total amount of energy utilized by the prosumers thanks to the anonymous billing via the smart meter. However, to preserve safety, some information about the prosumers needs to be public to allow checking of the offers to ensure that they are safe, or limit the resources available to them.

In our design, we assume that the consumption ($ECL$) and production ($EPL$) limits of each prosumer are public information, as well as which feeder a prosumer is on. The group safety constraints $C_{g}^{int}$ and $C_{g}^{ext}$ are also public. Recall that the smart contract ensures that no prosumer can withdraw more assets than the specified limits, and that any offer which violates the recorded safety constraints will be rejected. As a result, the only way to violate the safety requirements is if the asset limits or safety constraints are set incorrectly, which is not allowed by our design. However, as we will show below it is possible to improve privacy by choosing a conservative safety constraint for a group or a conservative limit on the maximum assets a prosumer can withdraw, which impacts the trading efficiency.  Consider the following example for illustration.
%to improve privacy by compromising efficiency of the system by choosing a conservative safety constraint for a group or a conservative limit on the maximum assets a prosumer can withdraw. 
%Consider the following example: 

%However, we can compromise efficiency of the system by choosing a conservative safety constraint for a group or a conservative limit on the maximum assets a prosumer can withdraw. Recall that the smart contract ensures that no prosumer can withdraw more assets than the specified limits, and that any offer than violates the recorded safety constraints will be rejected. The only way then to violate the safety requirements is if the asset limits or safety constraints are set incorrectly. 

 %To understand the impact on efficiency

%We have assumed that the consumption ($ECL$) and production ($EPL$) limits of each prosumer are public information, as well as which feeder a prosumer is on. The group safety constraints $C_{g}^{int}$ and $C_{g}^{ext}$ are also public. Given this information, consider the following example: 

 \begin{example}
    Given a group $g$ with an internal constraint of $C_{g}^{int}=30$ with prosumer $p1$ with $EPL=10$ and $p2$ with $EPL=20$. Assume that the prosumers in this group have anonymized their assets. If the total assets traded by the group ---call it $T_{f}$--- is below $10$ there is no way to definitively say that either prosumer is trading. If the assets traded by $f$ exceeds $10$ then we know that $p2$ is trading at least $T_{f} - 10$ since $p1$ can only produce $10$. If $T_{f} > 20$ then we know that $p1$ is trading at least $T_{f} - 20$. If $T_{f} = 30$ or $0$ then we know the full state of the feeder, either both are at maximum capacity or are off. To improve anonymity, the feeder as a whole should not trade more that $10$. This however reduces trading efficiency considerably. However if both prosumers have an $EPL=15$ then anonymity is improved until trading exceeds $15$. Thus it is important to select the constraints carefully. We discuss this in Section~\ref{sec:tradeoff}.
\end{example}{}

% \textcolor{blue}{PULLED FROM PETRA. NEED TO ADAPT.
% Due to our use of communication anonymity and mixing services, members of a microgrid can observe only the amount of assets withdrawn by a prosumer from its smart meter. Since all trading transactions are anonymous, they do not reveal the actual amount of assets traded by the prosumer. If a prosumer has not traded away all of its assets, then it can also anonymously deposit the remainder to a random address that was freshly generated by its smart meter. Even if a prosumer does not wish to trade, it should always withdraw, mix, and deposit the same amount of assets.  Otherwise, the lack (or varying amount) of withdrawal would leak information.As for the DSO, it receives the same information from the smart meter as in a non-transactive smart grid (i.e., amount of energy produced and consumed). Since trading is anonymous,the DSO learns only the financial balance of the prosumer,which is necessary for billing.  However, we can provide an even higher-level of privacy.  In particular, since price policies are recorded on the ledger (which the smart meters may read), each prosumer’s smart meter may calculate and send the prosumer’s monthly bill to the DSO, without revealing the prosumer’s energy consumption or production.  Meanwhile,the DSO can still obtain detailed load information (including predictions) for the microgrid from the bid storage and the trades recorded on the ledger.}

\subsection{Tradeoff between Privacy and Efficiency}
   \label{sec:tradeoff}
% \iAron{Note to self: think about separating safety and privacy.}
Note that safety of the system is a primary requirement and we cannot compromise it.
% feeders are groups. Safety is easy here just check feeder constraints. Forming groups must be done carefully and setting new constraints must be done carefully to ensure that trading is still safe while minimizing the lost trading potential.
%basic feeder constraints are given. Groups need new constraints. How you select those constraints impacts how much you can trade. 

\subsubsection{Selecting Group Constraints}
When the group is limited to a single feeder, the safety constraints are simply the feeder's constraints. However, in a group, once the accounts are anonymous, we cannot tell which feeder they belong to. Thus, to preserve safety, the constraints need to be modified. To do this, the set of feeders are transformed into a group by treating all the prosumers in those feeders as though they were on a common feeder. Since the offers are anonymous at the group-level, the system can treat the group as a single feeder with two prosumers: one which posts production offers and one which post consumption offers (see Fig. \ref{fig:feeder-conversion}). 

\begin{figure}
    \centering
    \begin{tikzpicture}[scale=.75,x=1cm,y=1cm,every text node part/.style={align=center},every node/.style={scale=.75}]

    \node(ss) at (0,0) {};
    
    \node(f1) at (.5,1) {};
    \node(f1end) at (3,1) {Feeder 1 \\ $C_f=10$};
    \node(f2) at (.5,-1) {};
    \node(f2end) at (3,-1) {Feeder 2 \\ $C_f=15$};
    
    \node(start) at (3.5,0){};
    \node(end) at (4.5,0){};
    \node(arr2start) at (8.5,0){};
    \node(arr2end) at (9.5,0){};
    
    \node(p1) at (1,1.5) {b1};
    \node(p2) at (1.5,.5) {s2};
    \node(p3) at (2,1.5) {b3};
    \node(p4) at (1,-.5) {b4};
    \node(p5) at (1.5,-1.5) {s5};
    \node(p6) at (2,-.5) {b6};
    
    \draw[](ss.center) -- (f1.center);
    \draw[](ss.center) -- (f2.center);
    \draw[](f1.center) -- (f1end);
    \draw[](f2.center) -- (f2end);
    
    \draw[]([shift={(.5,0)}]f1.center) -| (p1);
    \draw[]([shift={(1,0)}]f1.center) -| (p2);
    \draw[]([shift={(1.5,0)}]f1.center) -| (p3);
    
    \draw[]([shift={(.5,0)}]f2.center) -| (p4);
    \draw[]([shift={(1,0)}]f2.center) -| (p5);
    \draw[]([shift={(1.5,0)}]f2.center) -| (p6);
    
    \draw[->,ultra thick](start)--(end);

    \node(fs) at (5,0) {};
    \node(fp1) at (5.25,.5) {b1};
    \node(fp2) at (5.5,-.5) {s2};
    \node(fp3) at (5.75,.5) {b3};
    \node(fp4) at (6.25,.5) {b4};
    \node(fp5) at (6.5,-.5) {s5};
    \node(fp6) at (6.75,.5) {b6};
    \node(fsend) at (7.75,0){Group \\ $C_g$};
    
    \draw[](fs) -- (fsend);
    
    \draw[]([shift={(.25,0)}]fs.center) -| (fp1);
    \draw[]([shift={(.5,0)}]fs.center) -| (fp2);
    \draw[]([shift={(.75,0)}]fs.center) -| (fp3);
    
    \draw[]([shift={(1.25,0)}]fs.center) -| (fp4);
    \draw[]([shift={(1.5,0)}]fs.center) -| (fp5);
    \draw[]([shift={(1.75,0)}]fs.center) -| (fp6);
    
    \draw[->,ultra thick](arr2start)--(arr2end);
    
    \node(gs) at (10,0) {};
    \node(gp1) at (10.5,.5) {$\calB_{g}$};
    \node(gp2) at (11,-.5) {$\calS_{g}$};
    \node(gend) at (12,0){Group \\ $C_g$};
    
    \draw[]([shift={(.5,0)}]gs.center) -| (gp1);
    \draw[]([shift={(1,0)}]gs.center) -| (gp2);
    
    \draw[](gs)--(gend);

\end{tikzpicture}
    \caption{Feeder conversion diagram.}
    \vspace{-1em}
    \label{fig:feeder-conversion}
\end{figure}
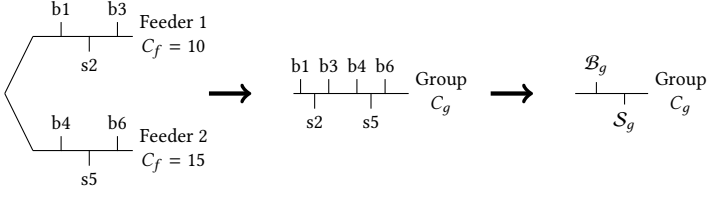

%\change{Groups give privacy through anonymity. However too much anonymity means that safety constraints cannot be enforced and agents are not held accountable. The smart contract ensures that no prosumer can withdraw more assets than the specified limits, and that any offer than violates the recorded safety constraints will be rejected. The only way then to violate the safety requirements is if the asset limits or safety constraints are set incorrectly.}{} 
Now we discuss how the group constraints can be set to ensure safety, and how much this costs in terms of trading efficiency.  We define this as the cost of privacy:
\begin{definition}\emph{Privacy Cost}:
% The cost of privacy is the amount of energy transfer that would have been satisfied by the trades, but was not in order to safely provide privacy.

In order to safely provide privacy, some amount of energy transfer (which would otherwise be satisfied by the trades) is lost. This is the cost of privacy.
\end{definition}

First, $EPL_{u}$ and $ECL_{u}$ are the same type of constraint representing a negative or positive flow of energy, so we will use $EL_{u}$ to represent both in our analysis, but in each case the equations refer to both. We also have the smart contract set $C_{g}^{ext}=C_{g}^{int}$ and refer to it as $C_g$ for now. %This assumption is reasonable when there is anonymity with multiple feeders it is impossible to tell if the trade is internal to a feeder or between feeders, and so we only care what the smallest constraint is. 

When setting the group safety limits, there are two cases to consider. We present them below. 
\begin{case}
    There is a set of prosumers in the group that is capable of exceeding the safety constraint of the feeder they are on. 
\end{case}{}
Assume a microgrid with feeders $\calF$ and groups $\calG$, wherein $EL$ for each prosumer and $C_{f}$ for each feeder can have any value. In order for this system to be safe, $C_{g}$ for every group $g$ must be:

    \begin{align}
        C_{g} \leq
            \min \left\{C_{f} \middle| f \in g \textnormal{ and } \sum_{u \in f}EL_{u} \geq C_{f}  \right\}
        \label{eq:groupLimit_int}
    \end{align}{}

\begin{proof}
    Assume Equation \eqref{eq:groupLimit_int} is false, and that the system is safe. This means $\exists C_{f} < \sum_{u \in f} EL_{u}$ and $C_{g} > C_{f}$. Let  $\sum_{u \in f} EL_{u} = C_{g}$--then the prosumers in $f$ can trade $EL$ assets. This exceeds the feeder safety limit, and the system is not safe. Equation \eqref{eq:groupLimit_int} must therefore be true. 
\end{proof}
Thus, the best value for the group constraint is when Equation \eqref{eq:groupLimit_int} is equality. This means that the group as a whole can at most produce the same amount as the single smallest of its internal feeders. The cost in this case is:
\begin{equation}
    cost = \min \left\{ \sum_{\forall s\in \calS_{g}}E_{s}, \sum_{\forall b\in \calB_{g}}E_{b} \right\} - \min \left\{ \sum_{\forall s\in \calS_{g}}E_{s}, \sum_{\forall b\in \calB_{g}}E_{b}, C_{g} \right\} .
\end{equation}
Simply, the cost is the amount by which the potential trades exceed the safety constraint.

\begin{case}
    No set of prosumers in any of the feeders in the group are capable of exceeding their feeders' safety constraint.
\end{case}{}

Given a microgrid with feeders $\calF$ and groups $\calG$ where $C_{f}$ can have any value and 
\begin{equation}
    \forall_{g} \forall_{f\in g} \sum_{u\in f} EL_{u} \leq C_{f} ,
    \label{eq:groupLimit_sum}
\end{equation}
group constraint should be set as $C_{g} = \sum_{f \in g} C_{f}$ to maximize trading, and trades can be done safely.
% then $C_{g} = \sum_{f \in g}\sum_{p \in f} EL_{p}$ can be traded safely.

\begin{proof}
    Assume a microgrid is not safe and Equation~\eqref{eq:groupLimit_sum} is true. Then, $\exists f$ such that $\sum_{u\in f} EL_{u} > C_{f}$. But, Equation~\eqref{eq:groupLimit_sum} says this is not allowed. So, the system is safe. 
\end{proof}

In this case, there is no cost to group privacy. The safety is provided by the the asset withdrawal limits rather than the group constraint. Note that case 1 can be converted to case 2 by modifying the prosumer limits so that the prosumers on a feeder cannot access assets that exceed their feeder's safety constraint. How the production and consumption limits are set can be negotiated by the DSO and the prosumers during installation. 

A consideration when converting from case 1 to case 2: since the prosumers have the capability to exceed  $C_{f}^{int}$, they also exceed $C_{f}^{ext}$. Now, if instead of being equal (as we assumed in the beginning), $C_{f}^{int} > C_{f}^{ext}$, then there is a cost associated with adding that feeder to a group, where the maximum cost is $C_{f}^{int} - C_{f}^{ext}$. 

This cost is incurred because trades internal to the feeder can't be distinguished from trades which are external, so the total amount traded by the feeder must not exceed $C_{f}^{ext}$. Otherwise, the external feeder over-current relay would trip. As a result, the prosumer must not be able to withdraw assets greater than their external feeder limit, resulting in the cost.

% if we group then we cannot use the higher internal limit since the trading must remain within the feeder or it will trip the over-current relay, unless $C_{f}^{int} = C_{f}^{ext}$ but we said that the internal limit was greater so there is a cost. 

\iflong
Within a group the accounts are allowed to transfer assets between each other. This is required so that the prosumers can deposit their unused assets back into an anonymous account owned by their smart meter. This means that a prosumer in one feeder can withdraw assets and transfer them to an account in another feeder, if they are colluding for example. Then the account in the second feeder can post offers that it cannot satisfy. However this is not a problem because the first prosumer will still be accountable for the missing assets. 

The fact that prosumers can exchange assets between anonymous accounts also means that prosumers on a feeder could transfer accounts between each other if one did not want to use its assets, then they would transfer assets back after the trading so the prosumer that did not want to use its assets could account for them. This is something we may explore in a future work.
\fi

\subsubsection{Insights on Grouping} 

Based on the analysis of the effects of privacy on efficiency, the best strategy is to limit the trading assets of the prosumers such that they remain less than the feeder constraints. This means that all feeders can be safely grouped. The cost of grouping feeders is the loss of flexibility in trading due to the rigid asset limits. The cost will be at most the feeder limit minus the prosumer asset limit, if that prosumer has the capacity to reach the feeder limit, and if no other prosumers in its feeder are trading. This could be mitigated by an additional mixing and trading step within the feeder, but we have not examined this possibility in detail. 
There is a second criterion that may influence grouping decisions. There is information leakage, and at the extremes (max load, zero load) anonymity ceases to exist. We assume that generally this will not be the case, and the odds of that occurring diminish if there are many feeders in the group.  Information leakage can be reduced by setting all the asset limits to the same value for all prosumers. The maximum system cost of this is the difference between the feeder limit and the sum of the prosumer limits. To reduce information leakage, groups should consist of feeders with similar limits.

   % \subsection{Privacy}\label{sec:privacy}
        
  %  \input{tmp/privacy.tex}

%\end{new}

%-----------------------------------------------------------
% TESTBED
%-----------------------------------------------------------

\begin{new}

    \section{Experimental Evaluation}
    \label{sec:experimental}
    
    In this section, we present a simulation testbed that we developed for evaluating \platform, as well as our initial results illustrating its effectiveness in reducing the load on the bulk power grid. 
    
    The system to demonstrate the simulation platform has three major parts. First, the simulation is controlled by a Python agent that sends simulation status and receives actuation commands from TRANSAX. Second, we use GridLAB-D as a discrete-time distribution network simulator. Third, messages and time steps between the Python agent and GridLAB-D are coordinated by the Framework for Network Co-Simulation\footnote{https://github.com/FNCS/fncs}.
    
    \subsection{Testbed}
    
    \subsubsection{Distribution System}
    
    \begin{figure}
        \centering
        \begin{minipage}[b]{0.4\textwidth}
            \centering
            \begin{tikzpicture}[
font=\tiny,
% scale=.8,
% every text node part/.style={align=center},
% every node/.style={scale=.6},
% relay/.style={rectangle, fill=black, minimum width=.05cm, minimum height=.5cm}
]

    \draw(0,0)--(0,2);
    \draw(-2.25,1)--(0,1);
    \draw(0,0.4)--(2,0.4);
    \draw(0,1.6)--(2,1.6);

    \draw(-2,1) -- (-2,1.25)node[anchor=south]{$101$};
    \draw(-1.8,1) --  (-1.8,.75)node[anchor=north]{$102$};
    \draw(-1.3,1) -- (-1.3,.75) node[anchor=north]{$103$};
    \draw(-0.8,1) -- (-0.8,.75) node[anchor=north]{$104$};
    \draw(-.6,1)--(-0.6,1.25) node[anchor=south]{$105$};
    
    \draw(0.4,1.6)--(0.4,1.35) node[anchor=north]{$201$};
    \draw(.9,1.6)--(.9,1.35) node[anchor=north]{$202$};
    \draw(1.15,1.6)--(1.15,1.85) node[anchor=south]{$203$};
    \draw(1.4,1.6)--(1.4,1.35) node[anchor=north]{$204$};
    \draw(1.65,1.6)--(1.65,1.85) node[anchor=south]{$205$};

    \draw(0.4,0.4)--(0.4,.65) node[anchor=south]{$301$};
    \draw(0.65,0.4)--(0.65,0.15) node[anchor=north]{$302$};
    \draw(1.15,0.4)--(1.15,0.15) node[anchor=north]{$303$};
    \draw(1.65,0.4)--(1.65,0.15) node[anchor=north]{$304$};
    \draw(1.9,0.4)--(1.9,.65) node[anchor=south]{$305$};
    
\end{tikzpicture}
            \caption{Topology of the simulated distribution network.}
            \label{fig:feeder}
        \end{minipage}
        \hspace{1em}
        \begin{minipage}[b]{0.5\textwidth}
            \begin{tikzpicture}[x=1cm, y=1cm, font=\tiny,
  Component/.style={fill=white, draw, align=center, rounded corners=0.1cm},
  Connection/.style={<->, >=stealth, shorten <=0.05cm, shorten >=0.05cm},
  call/.style={->, >=stealth, shorten <=0.05cm, shorten >=0.05cm},
  event/.style={<-, >=stealth, shorten <=0.05cm, shorten >=0.05cm}]

\node[Component, minimum height=2cm] (gl) at (0,0){GridLAB-D};
\node[Component, minimum height=2cm] (py) at (3,0){Python \\ Agent};

\node [Component,minimum height=2cm] (tx) at (6,0) {TRANSAX};

\draw[call]([shift={(0,.75)}]gl.east) to node [align=center, yshift=0.013cm] { Battery Charge \\ Solar Generation} ([shift={(0,.75)}]py.west);

\draw[call]([shift={(0,-.75cm)}]py.west) to node [align=center, yshift=0.11cm] { Battery Power Out} ([shift={(0,-.75cm)}]gl.east);

%----------------------------------------------------------------

\draw[call, dashed]([shift={(0,.6cm)}]tx.west) to node [align=center, yshift=0.12cm] { (1) Charge Request} ([shift={(0,.6cm)}]py.east);

\draw[call]([shift={(0,.25cm)}]py.east) to node [align=center, yshift=0.12cm] { (2) Battery Charge} ([shift={(0,.25cm)}]tx.west);

\draw[call, dashed]([shift={(0,-.25cm)}]tx.west) to node [align=center, yshift=0.12cm] { (3) Energy to Trade} ([shift={(0,-.25cm)}]py.east);

\draw[call]([shift={(0,-.6cm)}]py.east) to node [align=center, yshift=0.12cm] { (4) Step Request} ([shift={(0,-.6cm)}]tx.west);

\end{tikzpicture}%
            \caption{Message structure between simulator and TRANSAX.}
            \label{fig:testbed-msg}
        \end{minipage}
    \end{figure}

    The distribution system is modeled using GridLAB-D and is simulated on an x86 computer with and i7 processor and 24GB of RAM. The distribution topology (see Fig.~\ref{fig:feeder}) consists of a single substation feeding three main overhead lines that are connected to prosumers. The lines below the main lines represent prosumers that are modeled as houses with battery and solar panels can either consume or produce energy depending on the net output of the solar panels and battery, and those above represent prosumers that have loads and only consume. \iflong{The total number of nodes can be scaled into the hundreds of prosumers.}\fi For the demonstration discussed in this paper, the simulation was built with 9 producer nodes and 6 consumer nodes. The control logic of each prosumer is executed on single board computer (specifically the BeagleBone Black \footnote{https://beagleboard.org/black}). 
    
  %  \iflong{
    \subsubsection{Simulation Time Synchronization}    
    The Python agent is the interface that relays data between the GridLAB-D simulation and \platform, as well as synchronizes time between GridLAB-D's variable time-step solver and \platform's matching solver. The solution frequency is a system parameter. In this demonstration, \platform posted trades for each 15-minute interval of logical~time. 
    
    The most important feature of this demonstration is its methodology for synchronizing time between GridLAB-D and \platform. As noted above, GridLAB-D and \platform are time synchronized through the Python Agent. The Python agent forces GridLAB-D's variable time step solver to pause at each logical 15-minute interval. The Python agent waits until it receives a step request to continue from \platform. While GridLAB-D is paused, the Python agent and \platform send all the necessary messages for \platform to find and post finalized trades. 
    After 2 minutes of real-time, \platform sends a step request to the Python agent, and the Python agent actuates its control parameters for the next interval and commands GridLAB-D to simulate the next 15 minutes of logical time. This process recurs iteratively for the duration of the simulation's logical time. The time synchronization strategy outlined is scalable to any desired time period for the \platform solver. The strategy provides freedom to run experiments such as how the solver's time period effects the energy traded, the stability of the finalized trades, and the computational complexity.
    %}
 %   \fi
    
    \subsubsection{System Message Structure}
    
    The general message structure between GridLAB-D, the Python agent, and \platform is shown in 
    % \Aron{There is a lot of whitespace in this figure, can we shrink it?}\Scott{We could, but its the same size as the figure its next to, so it won't save any space.}
    Fig. \ref{fig:testbed-msg}. While GridLAB-D is paused, \platform agents request charge status for their batteries in the GridLAB-D simulation. They use this data, along with their predicted energy usage, to create a bid which is sent to \platform. \platform agents send the finalized trades back to the Python agent. The Python agent sets each simulated node's power output for the next interval based on the finalized trades from \platform by modifying GridLAB-D system parameters. In this demonstration, the Python agent only meets the finalized trades by modifying battery power outputs. However, the Python agent has control over all the dynamically modifiable parameters in GridLAB-D. As a consequence, future demonstrations could incorporate more control parameters such as curtailments to solar output or curtailments to energy used by pure consumers. 
    
    \subsubsection{Setup}
    The experiment setup to test \platform and the simulation contained 9 producers and 6 consumers for a total of 15 nodes in the distribution network. The simulation ran in logical time from 08:00 to 20:00 for a total of 12 hours. The experiment was run in two scenarios. First, the simulation was run without a link to \platform. In this scenario, the batteries were not set to output any power to demonstrate the behavior of the system without a transactive energy system. Second, the simulation was run in conjunction with \platform to demonstrate the efficacy of the transactive energy solver. During our experiments, we speed up the simulation by letting the real-time length of the time interval be $\hat{\Delta} < \Delta$ where $\hat{\Delta}=2 min$ and $\Delta=15 min$. Note that $\hat{\Delta}$ is the amount of real time passed in the simulation before proceeding to the next interval. This allows us to speed  up the experiments without compromising our results since running the system slower would be easier. 
    
    % With this setup since it was a private network once a block was mined the data was disseminated quickly and trades were delivered within seconds of finalization, so in this system we found that using $T_{clear}=1$ was effective. $T_{h}$ was set to similarly, the number of offers being posted was fairly small, 

    \begin{figure}
        \centering
        \begin{minipage}[b]{.49\textwidth}
            \pgfplotstableread[col sep=comma]{figs/horizon.csv}\horizon

\begin{tikzpicture}[scale=.7]

\begin{axis}[
    height=5cm,width=9cm, 
    axis y line*=left,
    ylabel={Memory Usage {[MB]}},
    grid
    ]
    \addplot table [x=Horizon, y=Mem, col sep=comma] {\horizon};
    \label{mem_plot}
\end{axis}

\begin{axis} [
    height=5cm,width=9cm, 
    xlabel={Solve horizon $T_{h}$}, 
    axis y line*=right,
    ylabel={Energy {[kW]}}, 
    legend style={at={(0.5,-0.15)},anchor=north,legend columns=-1 }
    ]
    \addplot[red,mark=*] table [x=Horizon, y=EnergyTraded, col sep=comma] {\horizon};
    \addlegendentry{Energy Traded {[kW]}}
    \addlegendimage{/pgfplots/refstyle=mem_plot}\addlegendentry{Memory {[MB]}}
\end{axis}
\end{tikzpicture}
            \vspace{-.5cm}
            \caption{Memory consumption and energy traded during a single interval of the simulation for various values of $T_{h}$ using the CPLEX solver.}
            \label{fig:loahNrgMEM}
        \end{minipage}
        \vspace{.5cm}
        \hfill
        \begin{minipage}[b]{.49\textwidth}
            \pgfplotstableread[col sep=comma]{figs/solveTime.csv}\solve

\begin{tikzpicture}[scale=.7]

\begin{axis}[height=5cm,width=9cm,grid, ylabel={Time {[ms]}}, date coordinates in=x, xticklabel= \hour:\minute]
%  [height=5cm,width=9cm,grid,axis y line*=left,ylabel=Time{[ms]}, date coordinates in=x, xticklabel= \hour:\minute,legend style={at={(0.5,-0.15)},anchor=north,legend columns=-1 }]
    
    \addplot[SkyBlue,mark options={scale=.75},mark=square*] table [x=interval_time, y expr=\thisrow{solveTime}*1000, col sep=comma] {\solve};
    
    \addplot[SeaGreen,mark options={scale=.75},mark=square*] table [x=interval_time, y expr=\thisrow{on_solve_time}*1000, col sep=comma] {\solve};

    \legend{Solve, Solve and Submit }
    \label{fig:solveTime}

\end{axis}

\end{tikzpicture}

% \begin{tikzpicture}[scale=.75]

% \begin{axis}[axis x line*=top, date coordinates in=x]
%     \addplot[] table {x="interval_time"}
% \end{axis}

% \end{tikzpicture}

% \begin{tikzpicture}[scale=.75]
% \begin{axis}[date coordinates in=x, xticklabel= \hour:\minute]
%     \addplot[] table [x="Time"] {\solve};
% \end{axis}
% \end{tikzpicture}

% \begin{tikzpicture}[scale=.75]
% \begin{axis}[]
%     \addplot[] table [x="solveTime"] {\solve};
% \end{axis}
% \end{tikzpicture}
            \caption{\textit{Solve} time is how long it took the solver to find a solution to the energy trading problem. \textit{Solve and Submit} time is how long to took to find the solution and submit it to the smart contract.}
            \label{fig:solve}
        \end{minipage}
        \begin{minipage}[b]{.49\textwidth}
            \pgfplotstableread[col sep=comma]{figs/substn.csv}\substn

\begin{tikzpicture}[scale=.7]
    \begin{axis}[height=5cm,width=9cm,ylabel={Real Power {[kW]}}, grid, 
    date coordinates in=x, xticklabel= \hour:\minute,
    legend style={at={(0.5,-0.15)},anchor=north,legend columns=-1 }]
        \addplot[thick,Dandelion] table [x=time, y expr=\thisrow{Original}/1000, col sep=comma] {\substn};
        \addplot[thick,Cerulean] table [x=time, y expr=\thisrow{WithTransax}/1000, col sep=comma] {\substn};
        \legend{Original, With TRANSAX}
    \end{axis}

\end{tikzpicture}
            \caption{Substation load with and without TRANSAX.}
            \label{fig:substn}
        \end{minipage}
        \hfill
        \begin{minipage}[b]{.49\textwidth}
            \pgfplotstableread[col sep=semicolon]{figs/chargeLvl.csv}\charge

\begin{tikzpicture}[scale=.7]

\begin{axis}[height=5cm,width=9cm,grid,ylabel=Charge Level{[\%]}, date coordinates in=x, xticklabel= \hour:\minute,legend style={at={(0.5,-0.15)},anchor=north,legend columns=-1 }]
    \addplot[Purple,mark options={scale=.75},mark=square*] table [x="Time", y expr=\thisrow{"charge.sum"}/1260, col sep=semicolon] {\charge};
\end{axis}

\end{tikzpicture}
            \caption{Average battery charge level.}
            \vspace{.5cm}
            \label{fig:chargeLvl}
        \end{minipage}
        \begin{minipage}[b]{.49\textwidth}
            \pgfplotstableread[col sep=comma]{figs/powerTraded.csv}\pTrade

\begin{tikzpicture}[scale=.7]

\begin{axis}[height=5cm,width=9cm, grid, ylabel={Power {[kW]}}, date coordinates in=x, xticklabel = \hour:\minute,legend style={at={(0.5,-0.15)},anchor=north,legend columns=-1 }]
    \addplot[Cerulean,mark options={scale=.75},mark=*] table [x=Time, y expr=\thisrow{TotalPowerTraded}/1000, col sep=comma] {\pTrade};
    \addplot[Green,mark options={scale=.75}, mark=square*] table [x=Time, y expr=\thisrow{selling}/1000, col sep=comma] {\pTrade};
    \addplot[Maroon,mark options={scale=.75},mark=triangle*] table [x=Time, y expr=\thisrow{buying}/1000, col sep=comma] {\pTrade};
    \legend{Trades, Selling, Buying}
\end{axis}

\end{tikzpicture}
            \caption{Green: sum of all production offers for each interval. Red: negative sum of all consumption offers for each interval. Blue: sum of all energy traded in each interval, whose maximum value is the minimum of the production and consumption offers.}
            \label{fig:powerTraded}
        \end{minipage}
        \hfill
        \begin{minipage}[b]{.49\textwidth}
            \pgfplotstableread[col sep=semicolon]{figs/solar.csv}\sTrade

\begin{tikzpicture}[scale=.7]

\begin{axis}[height=5cm, width=9cm, grid, ylabel={Power {[kW]}}, date coordinates in=x, xticklabel= \hour:\minute,legend style={at={(0.5,-0.15)},anchor=north,legend columns=-1 }]
    \addplot[Cerulean,mark options={scale=.75},mark=*] table [x="Time", y expr=\thisrow{"TotalEnergyTraded.sum"}/1000, col sep=semicolon] {\sTrade};
    \addplot[Purple,mark options={scale=.75},mark=square*] table [x="Time", y expr=\thisrow{"battActual.sum"}/1000, col sep=semicolon] {\sTrade};
    \addplot[Dandelion,mark options={scale=.75},mark=x] table [x="Time", y expr=\thisrow{"solarActual.sum"}/1000, col sep=semicolon] {\sTrade};
    \addplot[Orange,mark options={scale=.75},mark=*] table [x="Time", y expr=\thisrow{"totalActual.sum"}/1000, col sep=semicolon] {\sTrade};
    \legend{Trades, Battery, Solar, Actual}
\end{axis}

\end{tikzpicture}
            \caption{Yellow: simulated solar profile. Purple: simulated battery charge level. Orange: simulated energy traded. Blue: total energy trades recorded in the market.}
            \vspace{.5cm}
            \label{fig:soloar}
        \end{minipage}
    \end{figure}

    \subsection{Results}
    \label{sec:results}
    
    \subsubsection{Varying the time horizon}
    
    We ran multiple simulations of a microgrid varying the value of $T_h$ and selected a time interval during which we measured the memory usage and energy traded. In Fig. \ref{fig:loahNrgMEM}, we see that as the time horizon increases so does the memory usage and energy traded until $T_h=30$ at which point there is not additional gain to energy traded. The time horizon also impacts the CPU utilization of the solver  (not shown). This demonstrates that we can improve solver performance and still obtain quality solutions. 
    
    \subsubsection{Trading Impact}
    
    The simulation was first run without battery power output and without any control by \platform. This output was used to generate an energy profile for each prosumer for each interval. Then. the simulation was repeated with the prosumers submitting offers matching their energy profile to the \platform system to test the effect that batteries can have to mitigate load and utilize solar overproduction. For this reason, this test represents an ideal scenario with accurate bids for each 15 minute interval. Fig. \ref{fig:substn} shows the comparison of substation loads with and without \platform. The horizontal axis is the simulated time since the start of the simulation. The vertical axis shows the power load on the substation, negative values signify that prosumer generation exceeds their loads. Without \platform the system solar generation begins to outproduce the system load within the first interval. The solar reaches its peak production around 12:45. Finally, at 16:45, the load becomes greater than the solar production and the substation load is positive. The inclusion of \platform dramatically reduces the need for the substation backup. From 8:00 to 16:45, the overproduction of solar meant that the batteries on the system were charging, which mitigated the negative load on the substation. After 16:45, the batteries on the system discharged and mitigated the positive load on the system. Fig. \ref{fig:chargeLvl} shows the average battery charge level across all 9 batteries. At the end of the simulation (20:00), the average battery only had a 25\% charge. This means that if the simulation were to go further into the night, there would not have been enough battery capacity to power the house loads for the entirety of the night.
    
    % Fig. \ref{fig:chargeLvl} shows the average battery charge level across all 9 batteries in the simulated distribution network. This plot reiterates the results seen in Fig. \ref{fig:substn}. The batteries started the simulation with a 20\% charge and began charging around 9:15. The batteries were charged to mitigate overproduction until interval 16:45, when they started to discharge due to the load on the system being higher than generation. At the end of the simulation (20:00), the average battery only had a 25\% charge. This means that if the simulation were to go further into the night, there would not have been enough battery capacity to power the house loads for the entirety of the night. 
    
    Fig. \ref{fig:powerTraded} shows the total amount of energy offered for each interval, as well as the total amount of energy recorded in trades. In Fig. \ref{fig:soloar} we see that the trades recorded (Blue) are reasonably consistent with the measured load (orange) on the system, with one notable exception at 14:15. \iflong{The reason for these deviations is due to how we had the prosumers predicting their solar output. We go into more detail on this in the next section.}
    \else{The deviations occur because currently the prosumers assume that over an interval the solar input will remain constant, and this value is used when making offers. }

    Fig. \ref{fig:solve} shows the time required by the platform to find the optimal matching of a set of offers (green), as well as that time combined with the time required to submit that solution to the smart contract (blue). The majority of the time spent is due to the smart contract communications. 
    
    The results of the simulation with \platform are promising. \platform found energy trade solutions for each interval that resulted in an overall mitigation of substation load. The distribution was not completely independent of the substation feeder; there was still a need for a connection to the larger distribution grid.
    
    Finally, these results demonstrate that the developed simulation platform can effectively time synchronize and integrate with a transactive energy system like \platform. This simulation platform is flexible and reusable, and it can be used in future experiments to thoroughly investigate the performance of \platform for many different parameters.

    \iflong{
    \subsubsection{Extentions} 
    
%   In this demonstration, TRANSAX is not configured to consider the safety constraints of the system when determining trade solutions, and the inverters, transformers, and overhead lines of the simulation are all sized so that the system will still operate well even in a maximum load scenario. Future work for this testing platform can incorporate maximum current and power constraints into the TRANSAX solver such that it rejects any solutions that are unsafe for the simulated distribution network. Additionally, GridLAB-D can be configured with varying bottlenecks in the form of undersized inverters, transformers, or overhead lines to test whether the solver can still find valid solutions that do not overload the system, even if system components are undersized. 
    
    The python agent assumes constant solar output for the duration of an interval, so when it receives a finalized energy trade it computes the necessary battery output as the difference between the trade and the current solar value. If there is a change in the solar production or the prosumer introduces a load that is meant to be satisfied by its own power the battery output is not updated and will cause deviations from the value of the finalized energy trade.  An example of this can be seen in Fig. \ref{fig:substn}, from interval 21 to 22. There was a shift in load during that interval, and the overall system load increased to a maximum of over 5kW, even though there was sufficient battery capacity to mitigate the load. Additionally, in this demonstration, the solver did not have the ability to curtail solar overproduction or curtail loads that were not approved in the finalized trades. 
    % As discussed, in this demonstration, the Python agent receives a finalized energy trade for each node at each interval and uses current solar energy output to dictate each node's battery output for the next interval. If a drastic change in load occurs at a node in the middle of a fifteen minute interval, then the battery will not compensate until the beginning of the next fifteen minute interval, and the node will likely not trade the same amount of energy which was approved by the solver in the finalized trade. An example of this can be seen in Fig. \ref{fig:substn}, from interval 21 to 22. There was a shift in load during that interval, and the overall system load increased to a maximum of over 5kW, even though there was sufficient battery capacity to mitigate the load. Additionally, in this demonstration, the solver did not have the ability to curtail solar overproduction or curtail loads that were not approved in the finalized trades. 
    The Python agent can be improved to have better prediction capabilities or monitor the load more frequently to be able to dynamically modify battery power output, curtail solar output, or curtail house energy consumption in the middle of an interval if an unexpected shift in load is identified which causes a node to stray from its finalized trade for that interval. 
    
    The demonstration in this paper consisted of 15 nodes. While a small simulation is useful to demonstrate the effects of the trades finalized by TRANSAX, it is not realistic to test how a system like TRANSAX would act in a large, real-world distribution network. The developed system, with GridLAB-D as its simulator backbone, supports larger distribution networks. The principles of time synchronization and communication between the simulator and TRANSAX are identical in a large simulated system, so future tests with larger systems require minimal input other than redeveloping the GridLAB-D model. With this in mind, future models and tests should be established to demonstrate how the topology and size of a distribution network affect the ability of TRANSAX to finalize usable trades for every node on network. } \fi

\end{new}

\section{Related Work}
\label{sec:related}

\subsection{Existing Deployments}
Existing deployments are limited and so far have not published results. The closest is W\"{o}rner et al.~\cite{worner_trading_2019}, who have developed an implementation of their peer-to-peer energy market and deployed it to a town in Switzerland. Their goal is to to gather empirical evidence to answer the question of what are the benefits of a blockchain system in the electricity use case. Their study will conclude at the end of 2019, at which time they will analyze the data to determine the performance of the blockchain system. To motivate the design of their system, they carried out a targeted literature review, and selected ``a double auction with discriminative pricing as the most suitable market mechanism for electricity exchange.'' 

\iflong{
\subsection{Transactive Energy Markets}

    \subsubsection{Active Energy Markets}
    Blockchain based energy trading has changed from a theoretical topic to working and practical solutions all over the world\cite{app9081561}. First working application of a blockchain energy market was created back in 2016 named Exergy~\cite{Orsini2019Approach}. Currently, Germany, France, Australia, and Switzerland are among countries which are using such energy markets.
    
    \begin{description}
        \item[France] SunChain~\cite{sunchain} is the first active blockchain energy market in France which uses Hyperledger as underlaying datastore for solar energy prosumers. It aims to provide billing and transaction tracking at low cost. I-NUK~\cite{inuk} is a similar platform which is currently being developed whose goal is to reduce the daily carbon footprint.
        \item[Germany] Tal.Markt~\cite{talmarkt} in Germany tries to create an open energy market for solar energy producers with capacity more than 30 kWh to sell their energy to local citizens.
        \item[Australia] Power~Ledger~\cite{PowerLedger} creates an automated settlement system which is trustless, transparent, and interoperable that helps energy retailers to exchange energy and funds.
        \item[Switzerland] PowerID~\cite{powerid} is a local P2P market in Switzerland which wants reduce local energy delivery costs.
    \end{description}
}
\fi
    
    \subsection{Buyer-Seller Matching}
    
    After prosumers presented their energy availabilities and demands in form of offers, these offers need to be matched. Researchers have proposed two approaches for this problem.
    
    \subsubsection{Stable Matching} Stable matching refers to matching of all possible buy and sell offers in a bipartite graph. Yucel \textit{et al.} proposed a homomorphic encryption-based position hiding method\cite{yucel2018privacy} which protects users' privacy from adversary matchers. Nunna \textit{et al.}~\cite{kumar2013multiagent} proposed the symmetrical allocation problem based on native auction algorithm to match buyers and sellers. \iflong{This algorithm is run periodically, and only one selling offer is matched in each round.}\fi PowerLedger~\cite{PowerLedger} uses another mechanism to match offers. Offers are broken into equal portions and matched together e.g., when a new consumer arrives, it will receive the equal allocation from the energy pool in the~area. \iflong{We have designed an automatic stable matching mechanism in \platform to ensure prosumers' privacy and system's safety.}\fi
    \subsubsection{Auction} 
    \label{sec:auction}
    Another approach to match buyers' offers to sellers' is to use auctioning approaches. Majumder \textit{et al.}\cite{majumder_efficient_2014} proposed a double auction mechanism before the era of blockchains where the controller doesn't need the users' data. As a result, the privacy of bidders and sellers will be preserved. 
    % Further, \cite{Vytelingum:2010:TAS:1838206.1838326,ramachandran2011intelligent,8534390} were among the first who used double auctioning in distributed grid transactions. 
    In the era of blockchains, Kang \textit{et al.}~\cite{7935397} and Guerrero~\cite{8513887} used double auctions to match parties and not goods in blockchains. To ensure integrity of results of matching, Wang~\cite{en10121971} proposed a multi-signed digital certificate. Khorsani \textit{et al.}~\cite{8282470} designed a greedy algorithm with the averaging auction mechanism to match buyers with higher price to sellers with lower prices. Zhao \textit{et al.}~\cite{en11092412} created a two phase auctioning algorithm to find the optimal pricing for bids\iflong{; In the first phase, bidders and sellers provide guiding price for the next intervals} \fi. Finally, Zhang \textit{et al.}~\cite{zhang20181} developed a non-cooperative auctioning game and used it to find the optimal solution for the matching problem using the Nash Equilibrium.
        
    % Although buyer-seller matching is a requirement for every energy market, it is not the goal of \platform to design or develop such algorithms. \platform assumes the pricing control comes from DSO and we only focus on energy matches. Integration of pricing and auction markets is a straight forward extension of our approach.
    % \subsection{Blockchain Problems (Low efficiency [62] and high transaction costs [63,64])}
    
    \subsection{Online Information Management Platforms}
    Information Management Platform are designed to collect, process, transmit, and analyze data. In this context, data collection usually happens at the edge because that is where edge devices with sensors are deployed to monitor surrounding environments. \platform does not suggest a specific data collection methodology. Rather, it follows an actor-driven design pattern where ``prosumer'' actors can integrate their own agents into \platform by using the provided APIs.  Another concern of these platforms is the cost of processing. Traditionally, this problem was solved using scalable cloud resources in-house~\cite{schmidt2014elastic}. However, \platform enables a decentralized ecosystem, where components of the platform can run directly on edge nodes, which is one of the reasons why we designed it to be asynchronous in nature.
    
    To an extent, the information architecture of \platform can be compared to dataflow engines, such as Storm~~\footnote{http://storm.apache.org/}, Spark~\footnote{http://spark.apache.org/}, and S4 \cite{neumeyer2010s4}. All of these existing dataflow engines use some form of a computation graph, comprising computation nodes and dataflow edges. These engines are designed for batch-processing and/or stream-processing high volumes of data in resource intensive nodes, and do not necessarily provide additional ``platform services'' like trust management or solver~nodes. 
    % \subsection{Correctness of Smart Contracts}
    
    \subsection{Grid Control and Stability}
    % \Aron{Does this really belong to the `Buyer-Seller Matching' subsection?}
    One integral part of smart-grids are the microgrid controllers which ensure stability and resiliency of the microgrid. They enable transition of the microgrid from grid-connected to islanded~\cite{LIDULA2011186,USTUN20114030} so that the failures in the grid do not cascade to other areas similar to the outage event back in 1999 Sao Paulo, Brazil~\cite{PhysRevE.70.035101}. Currently, most of microgrid controllers are centralized~\cite{KAUR2016338} which are vulnerable to cyber-threats and privacy issues. A large spectrum of cyber-threats are applicable on centralized microgrid controllers with single-point-of-failure ranging from attackers eavesdropping on channels between the controllable resource and the centralized controller to steal critical information of the users or network infrastructure, performing DDOS attacks on the centralized controller, or manipulation of demand via IoT (MadIoT) attacks\cite{236250} to injecting malware into the market operation system and manipulate settings, such as DLMP limits or clearing time interval similar to the notable cyber-attack against Ukrainian power systems in December 2015~\cite{lee2016analysis,zetter2016inside}.
    
    Due to these drawbacks of centralized grid controls, industry is transforming from centralized to decentralized\cite{6870484,7112129}. The aim of \platform is to create a decentralized transactive energy market which ensures privacy and security of users while maintaining stability and resiliency of the grid.
    % A centralized trading platform may be exposed to a variety of cyber-threats and privacy issues. Some attackers seek financial gain through network-based attacks, and they will manipulate the controllers to profit. Some attackers aim to disturb the operation of the TES. Similar to the notable cyber-attack against Ukrainian power systems in December 2015~\cite{lee2016analysis,zetter2016inside}, attackers can inject malware into the market operation system and manipulate settings, such as DLMP limits or clearing time interval. An attacker could use a malicious channel to eavesdrop on the power system and can also steal critical information. Through stealing critical information from both the market operator and prosumers, attackers could devise a sophisticated targeted attack. They can also conduct Denial-of-Service (DOS) attacks that aim to cause a lack of availability of information, updates, prices, and resources. In contrast to the market operation system, individual smart HVAC controllers do not necessarily employ strong security mechanisms. Therefore, compromise a large number of smart HVAC controllers either for financial gain or to damage the system by sending massively manipulated bids to the market operator.

\subsection{Security and Privacy}

    \subsubsection{Communication Security}
    First step to preserve users' privacy and anonymity in a distributed system is to provide communication privacy. Without this, an adversary can discern who is making a function call or sending a message over the network based on the sender's MAC address, IP address, or route to destination. Existing protocols for low-latency communication anonymity include onion routing~\cite{reed1998anonymous}, the similar garlic routing \cite{Liu2014EmpiricalMA}, STAC \cite{7986314}, and the decentralized Matrix protocol~\footnote{https://matrix.org/docs/spec/}. However, Murdoch and Danezis \cite{1425067} show that a low-cost traffic analysis is possible of the Tor-network, theoretically and experimentally. Communication security is an orthogonal research problem to \platform.
    % \Aron{We can make this shorter (just comment out some parts, may be useful later).} 
    \iflong{
    Traffic analyses are based on tracking the forwarding of the size of a data package between computers, for example, if computer A sends a package of exactly 42 bytes to computer J, who then sends a package of exactly 42 bytes to B, it can be easily deduced that A sent a package of unknown content to computer B. This is possible because of the distribution of metadata to all routers in the Tor-network~\cite{Hopper:2010:MAN:1698750.1698753}. In what is called a timing analysis attack, an attacker tries to find a correlation between the timing of messages moving through the network to gain information about user identities and their communications. Analyses have shown that these types of attacks can be very effective over a wide range of network parameters when specific defences are not employed~\cite{Levine2004,4797313}.  To counter timing analysis attacks, the I2P network bundles multiple messages together (principle of garlic routing) and renders it more difficult to analyse~\cite{Liu2014EmpiricalMA}. Schimmer, 2009, showed that the bandwidth opportunistic peer-selection and -profiling algorithm does not prioritize anonymity in favor of performance~\cite{peerProfiling:2009}. Herrmann and Grothoff, 2011, exposed a potential weakness in anonymous HTTP-hosting done over the I2P network \cite{Herrmann2011}. The arguably only practical attack against the I2P network was done against the directory, the netDb, by Egger \textit{et al.} \cite{Egger2013}. An improvement of the protocol, aimed at Egger \textit{et al.}'s attack was suggested by Timpanaro \textit{et al.}, 2015 \cite{Timpanaro2015}. }\fi

    \begin{petra}{}
    
    \subsubsection{Address Anonymity}
    Communication anonymity is necessary but not sufficient for anonymous trading, as the cryptographic objectives of authentication and legitimacy are not fulfilled. We suggest using cryptographic techniques from distributed ledgers, \textit{blockchains}, and cryptocurrencies. The most adopted one, Bitcoin allows for very simple digital cash spending but has serious privacy and anonymity flaws~\cite{Barber2012,Reid2013,apostolaki2017}. Additionally, Biryukov and Pustogarov, 2015, show that using Bitcoin over the Tor network opens a new attack surface~\cite{biryukov2015}. Solutions to the tracing and identification problems identified by these researchers have been proposed and implemented in alternative cryptocurrency protocols: mixing using ring signatures and zero-knowledge proofs~\cite{cryptonote, miers2013zerocoin}.
    
    A proposed improvement to standard ring signatures is the CryptoNote protocol, which prevents tracing assets back to their original owners by mixing together incoming transactions and outgoing transactions. This service hides the connections between the prosumers and the addresses. Mixing requires the possibility to create new wallets at will and the existence of a sufficient number of participants in the network. Monero is an example of a cryptocurrency that provides built-in mixing services by implementing the CryptoNote protocol~\cite{cryptoeprint:2015:1098}. There are however alternative implementations of mixing protocols such as CoinShuffle~\cite{ruffing2014coinshuffle} or Xim~\cite{bissias2014sybil}. A variant of ring signatures, group signatures, were first introduced by Chaum and van Heyst, 1991,~\cite{Chaum1991} and then built upon by Rivest \textit{et al.}, 2001~\cite{Rivest2001}. The basis for anonymity in the CryptoNote protocol, however, is a slightly modified version of the \textit{traceable ring signature} algorithm by Fukisaki and Suzuki, 2007~\cite{Fujisaki2007}. This allows a member of a group to send a transaction so that it is impossible for a receiver to know any more about the sender than that it came from a group member without the use of a central authority.
    
    Some newer cryptocurrencies, such as Zerocoin~\cite{miers2013zerocoin}, provide built-in mixing services, which are often based on cryptographic principles and proofs.
    
    \end{petra}

    \subsubsection{Smart Meters' Privacy}
    Most works discussing privacy look at it from the context of smart meters. McDaniel and McLaughlin discuss privacy concerns due to energy-usage profiling, which smart grids could potentially enable~\cite{mcdaniel2009security}. Efthymiou and Kalogridis describe a method for securely anonymizing frequent electrical metering data sent by a smart meter by using a third party escrow mechanism~\cite{efthymiou2010smart}. Tan et al. study privacy in a smart metering system from an information theoretic perspective in the presence of energy harvesting and storage units~\cite{tan2013increasing}. They show that energy harvesting provides increased privacy by diversifying the energy source, while a storage device can be used to increase both energy efficiency and privacy. However, transaction data from energy trading may provide more fine-grained information than smart meter based usage patterns~\cite{Privacy2017}.

\section{Conclusions}
\label{sec:conclusion}

\begin{comment}
\Aron{We should reference equations by numbers instead of LaTeX IDs! Since numbering may change, do this just before submission.}

\begin{lstlisting}[language=Solidity, linewidth=\columnwidth, label={lst:constraints}, caption=Smart Contract constraints]
// check if buyer and seller are matchable
require(time >= sellingOffers[sellerID].startTime);
require(time <= sellingOffers[sellerID].endTime);
require(time >= buyingOffers[buyerID].startTime);
require(time <= buyingOffers[buyerID].endTime);
// eq:constrEnergyProd        
require(solution.sellerProduction[sellerID] <= sellingOffers[sellerID].energy);
// eq:constrEnergyCons       
require(solution.buyerConsumption[buyerID] <= buyingOffers[buyerID].energy);
// eq:constrIntProd
require(solution.feederProduction[sellingFeeder][time] <= Cint);
// eq:constrIntCons
require(solution.feederConsumption[buyingFeeder][time] <= Cint);
// eq:constrExtProd
require(solution.feederProduction[sellingFeeder][time] - solution.feederConsumption[sellingFeeder][time] <= Cext);
// eq:constrExtCons
require(solution.feederConsumption[buyingFeeder][time] - solution.feederConsumption[buyingFeeder][time] <= Cext);
\end{lstlisting}
\end{comment}

In this paper, we described \platform, a decentralized platform for implementing energy exchange mechanisms in a microgrid setting. Building on top of blockchains, we obtained decentralized trust and consensus capabilities, which prevent malicious actors from tampering with the shared system state. We demonstrated that the system is safe from anonymous offers, satisfying the seemingly conflicting goals of safety and privacy. Using our hybrid-solver approach, which combines a smart-contract based validator with an external optimizer, we showed that we can clear offers securely, efficiently, and resiliently. We also described a simulation testbed that allows us to test market platforms on a simulated microgrid, and we presented results showing how the platform can reduce the load on the bulk grid. 

\subsection{Future Work}
\label{sec:future}
\iffuture{
In this work, we focused on satisfying the requirements of safety, efficiency, privacy, and resilience. We have not chosen a specific approach for \emph{setting the clearing prices} for the prosumers' trades since the economics of setting the clearing prices is an orthogonal problem. Friedman and Rust~\cite{friedman_double_2018} provide a survey of these mechanisms for governing trade, to which they refer as  market institutions. One of the most commonly used mechanisms is the double auction. Note that we cannot apply the double auction directly because of the different time-interval attributes that the offers may specify. Prior work has extended the double auction to allow for multiple attributes; however, they typically (e.g. \cite{baranwal_fair_2015}) require a function to combine the attributes into a single value, which is then used to order the offers. The difficulty of this approach is in identifying a meaningful function. A more straightforward approach is to perform the feasibility matching as we have presented, and then for each interval, use a double auction to set the clearing price for the matched offers. This approach provides a straightforward solution to the problem of setting clearing prices; however, it is not obvious whether it will preserve the properties that a simple double auction has, such as incentive compatibility. %as a simple double auction  we are not sure if this will impact the properties that a double auction normally provides, like incentive compatibility. 
We leave the investigation of these mechanisms and how they are impacted by privacy to future work. 

% In future work, we will study these mechanisms, especially trying to understand the implication of the strict privacy controls on the efficiency of the economic market.

% Therefore, we worked on establishing the platform fundamentals and trading protocols, while assuming that the reservation prices are already available to the prosumers. Given this, prosumers decide how much resource to offer resulting in a \textit{Bertrand competition} market~\cite{kreps1983quantity}, where we ensure that all matched trades are feasible. 

A second extension is to enable prosumers to \emph{update or cancel offers}. The current formulation can support updating offers as long as the updates do not invalidate previous solutions; for example, a selling offer can increase the amount of energy to be sold or augment the set of intervals in which energy could be produced. To support restrictive changes or cancelling offers, we would need to introduce a deadline for when offers could no longer be updated or cancelled. Solvers could then wait for this deadline, and start working only after the deadline.

Finally, we have not yet discussed how the solvers are \emph{incentivized to match the offers}. One option is to provide some reward for each solution that is finalized. Another is to expect the prosumers to submit solutions that are beneficial to themselves. These and other potential incentive mechanisms have yet to be evaluated.

}\fi

\bibliographystyle{ACM-Reference-Format}
\bibliography{references} 

\appendix

\end{document}
\endinput
%%
%% End of file `sample-acmlarge.tex'.